\documentclass[aps,physrev,reprint,a4paper,superscriptaddress,citeautoscript,floatfix]{revtex4-2}

\usepackage{amsmath,amsfonts,amssymb}
\usepackage{newtxtext,newtxmath}
\usepackage{mathrsfs}

\usepackage[pdftex]{graphicx}
\usepackage{microtype}
\usepackage{bm}

\usepackage{siunitx}

\usepackage[svgnames]{xcolor}
\usepackage[colorlinks=True,linkcolor=blue,citecolor=blue,urlcolor=blue,pdfstartview=FitH,bookmarks=False,pdfpagemode=UseNone]{hyperref}

\begin{document}

\title{Band modulations and topological transitions in a one-dimensional periodic bead-on-string chain}

\author{Haocong Pan}
\affiliation{School of Physics, Peking University, Beijing 100871, China}
\author{Wei Wang}
\affiliation{School of Physics, Peking University, Beijing 100871, China}
\author{Chunling Liu}
\affiliation{School of Physics, Peking University, Beijing 100871, China}
\email{clliu@pku.edu.cn}
\date{\today}

\begin{abstract}
We study band modulations and topological transitions in a one-dimensional periodic bead-on-string chain.
Using an exact transfer-matrix formulation of the wave equation with periodically modulated mass density, combined with numerical spectral searches and tabletop experiments, we characterize band gaps and localized midgap states.
We interpret these states by mapping the system to the Su-Schrieffer-Heeger (SSH) model and its low-energy (1+1)-dimensional Dirac theory.
This framework reveals that the robust states are topological solitons bound to boundaries or engineered domain walls in the Dirac mass.
Through this mapping, we provide an intuitive account of how band structure controls topological phase changes in mechanically realizable lattices.
\end{abstract}

\maketitle

\section{Introduction\label{sec:intro}}
Topological phases and their transitions are currently hot topics in condensed matter physics.
Unlike conventional phases described by local order parameters, many topological phases are distinguished by global invariants and robust boundary phenomena.
The transitions between distinct topological regimes generally require bulk gap closings and the reconfiguration of these invariants \cite{TKNN1982,Hatsugai1993,HasanKane2010,QiZhang2011}.
In one dimension, a minimal and illuminating example is the Su-Schrieffer-Heeger (SSH) model, a bipartite tight-binding chain with alternating hopping amplitudes.
By tuning the relative strength of intra-cell and inter-cell hoppings, the bulk gap can close and reopen, and the system undergoes a topological transition signalled by the appearance or disappearance of midgap edge-localized modes \cite{SuSchriefferHeeger1979,SuSchriefferHeeger1980}.
Near the gap-closing point, the lattice Hamiltonian admits a low-energy expansion that maps onto a (1+1)-dimensional Dirac equation with a mass term whose sign distinguishes the two phases.
Therefore, a spatial domain wall in this mass traps a localized mode via the Jackiw-Rebbi mechanism \cite{JackiwRebbi1976}.
This mechanism provides the continuum description of a topological soliton bound to a domain wall in the Dirac mass, a concept originally formulated in quantum field theory that now finds concrete realizations in condensed matter and classical wave systems.

The above quantum-mechanical ideas translate directly to wave-based platforms governed by classical field equations: photonic systems (Maxwell's equations), acoustic systems (elastic wave equations), and mechanical lattices (Newton's equations).
In all these cases, the linear wave equations admit a Schrödinger-like form, enabling the realization of topological band phenomena.
Experimental demonstrations in mechanical lattices have confirmed the generality of this wave-physics perspective \cite{KaneLubensky2014,SusstrunkHuber2015}.
Beyond these platforms, electric circuits provide a flexible platform for realizing topological states, where disorder can induce a topological Anderson insulator phase \cite{ZhangWuSongJiang2019}.
Similar disorder-driven topological transitions have also been observed in one-dimensional mechanical systems \cite{ShiKiorpelidisTheocharis2021}.
Furthermore, recent work shows that topological phases can arise from nontrivial structural units themselves, independent of perfect periodicity \cite{ZhangLiWang2025}.

In this work we focus on a mechanically simple but physically rich system: a tensioned string bearing a periodic sequence of beads, which we treat as point masses.
The spatially modulated linear mass density produces banded vibrational spectra and tunable band gaps.
When parameters or spatial order are varied, these gaps can close and reopen, leading to localized midgap states.
Employing an analytic, Kronig-Penney-type transfer-matrix formulation of the 1D wave equation \cite{KronigPenney1931}, together with numerical root-finding for finite chains and tabletop measurements of eigenfrequencies and mode profiles, we: (1) characterize how intra-cell masses and spacings tune the band structure and band gaps; (2) identify parameter regimes where dimerization induces SSH-like physics, leading to robust midgap edge modes \cite{SuSchriefferHeeger1979,SuSchriefferHeeger1980}; and (3) interpret these phenomena through effective SSH mappings and the (1+1)-dimensional Dirac mass-domain picture \cite{JackiwRebbi1976,RiceMele1982}, using bulk-edge correspondence as a guiding diagnostic \cite{Hatsugai1993}.
Our work aims to demonstrate how basic classical-mechanical principles combined with band-structure analysis can elucidate topological transitions in an experimentally accessible and conceptually clear mechanical system.

\section{Model and Approach\label{sec:model_and_approach}}
Before presenting our results, we outline the basic theoretical and experimental framework.
We formulate the wave equation for a string with point masses, recast it in transfer-matrix form to obtain band conditions, and describe how finite-chain spectra and mode shapes are extracted numerically.
We also introduce the simple diagnostics that guide our identification of edge states and the experimental setups used for measurements.

\subsection{Continuum wave equation with point masses\label{subsec:wave_eqn}}
We consider transverse vibrations $u(x,t)$ of a taut string under constant tension $T$.
The background linear mass density is $\rho_0$, and $N$ beads of masses $m_i$ are threaded at positions $x_i$, where $i=1,\cdots,N$ and all $x_i$ are measured from the left end.
The resulting linear mass density becomes
\begin{equation}
\rho(x)=\rho_0+\sum_{i=1}^N m_i\delta(x-x_i).
\end{equation}
The corresponding wave equation is
\begin{equation}
\frac{\partial^2 u}{\partial t^2}-\frac{T}{\rho(x)}\frac{\partial^2 u}{\partial x^2}=0.
\end{equation}
Adopting the harmonic ansatz $u(x,t)=U(x)e^{-i\omega t}$ yields
\begin{equation}
U''(x)+\omega^2\frac{\rho(x)}{T}U(x)=0.
\end{equation}
Between the beads, where $\rho(x)=\rho_0$, the solutions are combinations of $\cos(qx)$ and $\sin(qx)$ with $q=\omega\sqrt{\rho_0/T}$.

\subsection{Kronig-Penney-type transfer-matrix formulation\label{subsec:transfer_matrix}}
To connect solutions across different string segments, we define the local state vector $\Psi(x)=(U(x),\,T U'(x))^{\mathrm{T}}$.
The formalism is analogous to the Kronig-Penney model for electrons in a periodic delta-function potential, adapted here to a mechanical wave equation with point masses.

For a homogeneous segment of length $d$ with no beads, the propagation is described by
\begin{equation}
\mathbb{P}(d)=\begin{pmatrix}
\cos(qd) & \dfrac{\sin(qd)}{\sqrt{T\rho_0}}\\[6pt]
-\sqrt{T\rho_0}\,\sin(qd) & \cos(qd)
\end{pmatrix},
\quad q = \omega\sqrt{\rho_0/T},
\end{equation}
so that $\Psi(x+d)=\mathbb{P}(d)\Psi(x)$.

At a bead of mass $m$, the displacement remains continuous while the slope jumps due to the mass.
This boundary condition is encoded in the jump matrix
\begin{equation}
\mathbb{L}(m)=\begin{pmatrix}1 & 0\\[4pt]
-\dfrac{m\omega^2}{T} & 1
\end{pmatrix},
\end{equation}
which satisfies $\Psi(x^+)=\mathbb{L}(m)\Psi(x^-)$, where $x^{\pm}$ denote positions just to the right/left of the bead.

If a unit cell of length $a$ contains $n$ beads at positions $x_1,\dots,x_n$ with masses $m_1,\dots,m_n$, the transfer matrix across one full cell is the ordered product
\begin{equation}
\mathbb{M}_{\mathrm{cell}}(\omega)=\mathbb{P}(a-x_n)\mathbb{L}(m_n)\cdots\mathbb{P}(x_2-x_1)\mathbb{L}(m_1)\mathbb{P}(x_1).
\end{equation}
Imposing Bloch periodicity, $\Psi(x+a)=e^{ika}\Psi(x)$, leads to the condition
\begin{equation}
\cos(ka)=\frac{1}{2}\operatorname{Tr}\mathbb{M}_{\mathrm{cell}}(\omega).
\end{equation}
Consequently, the allowed frequency bands are those $\omega$ for which $|\operatorname{Tr}\mathbb{M}_{\mathrm{cell}}(\omega)|\leq 2$; outside this range the frequency lies in a band gap \cite{KronigPenney1931}.
This approach yields Kronig-Penney-type band diagrams that directly reveal the allowed and forbidden frequency ranges.

\subsection{Finite-chain quantization and mode reconstruction\label{subsec:finite_quant}}
For a finite chain of length $L$ with Dirichlet boundary conditions $U(0)=U(L)=0$, we construct the total transfer matrix $\mathbb{M}_{\mathrm{tot}}(\omega)$ by multiplying the propagation and jump matrices across the entire system.
Writing $\mathbb{M}_{\mathrm{tot}}(\omega)=\begin{pmatrix}M_{11}(\omega)&M_{12}(\omega)\\ M_{21}(\omega)&M_{22}(\omega)\end{pmatrix}$, the boundary conditions impose a quantization condition.
A convenient choice is $M_{12}(\omega)=0$, which enforces that a state starting with zero displacement and finite slope at $x=0$ also has zero displacement at $x=L$.
We solve this condition numerically by scanning the frequency axis and applying root-finding to locate the eigenfrequencies.

Once an eigenfrequency $\omega$ is found, the corresponding eigenmode $U(x)$ is reconstructed by forward propagation: starting from $\Psi(0)=(0,1)^{\mathrm{T}}$ at the left boundary, we apply the propagation and jump matrices step by step to obtain the displacement profile across the entire chain.

The spectrum of a finite periodic chain depends on how the unit-cell pattern aligns with the boundaries.
We parametrize this alignment by an offset $\tau$, defined as the distance from the left boundary to the leftmost bead.
Setting $\tau=0$ places a bead exactly at the left endpoint; other values shift the whole pattern along the chain while keeping the internal spacings fixed.
Varying $\tau$ changes the effective termination of the chain, which will prove useful below for exploring how edge modes depend on boundary details.

\subsection{Topological diagnosis via edge states\label{subsec:edge_diagnosis}}
In order to establish a direct, experimentally accessible link between bulk-band rearrangements and the appearance of robust boundary-localized modes, we adopt a pragmatic approach centered on edge states.
This strategy proceeds in two stages: first, we use the Kronig-Penney-type cell transfer-matrix calculation to locate gap closings and reopenings as parameters are varied; second, we examine the finite-chain spectra and mode profiles across these variations to identify modes that emerge within the reopened gaps.

A midgap mode is identified as a topological edge candidate if it satisfies two key requirements.
First, the mode must appear only after a bulk gap reopens and then persist smoothly within that gap under adiabatic changes of bulk parameters, maintaining spectral isolation from the bulk bands.
Second, the candidate must be robust against small, localized perturbations at the boundary, such as slight shifts of the end bead.
Trivial termination-induced states typically shift strongly or vanish under such perturbations, whereas a bulk-driven topological edge state remains in the gap and retains an exponentially localized profile.

\subsection{Experimental setup\label{subsec:experimental_setup}}

\begin{figure}
\includegraphics[width=\linewidth]{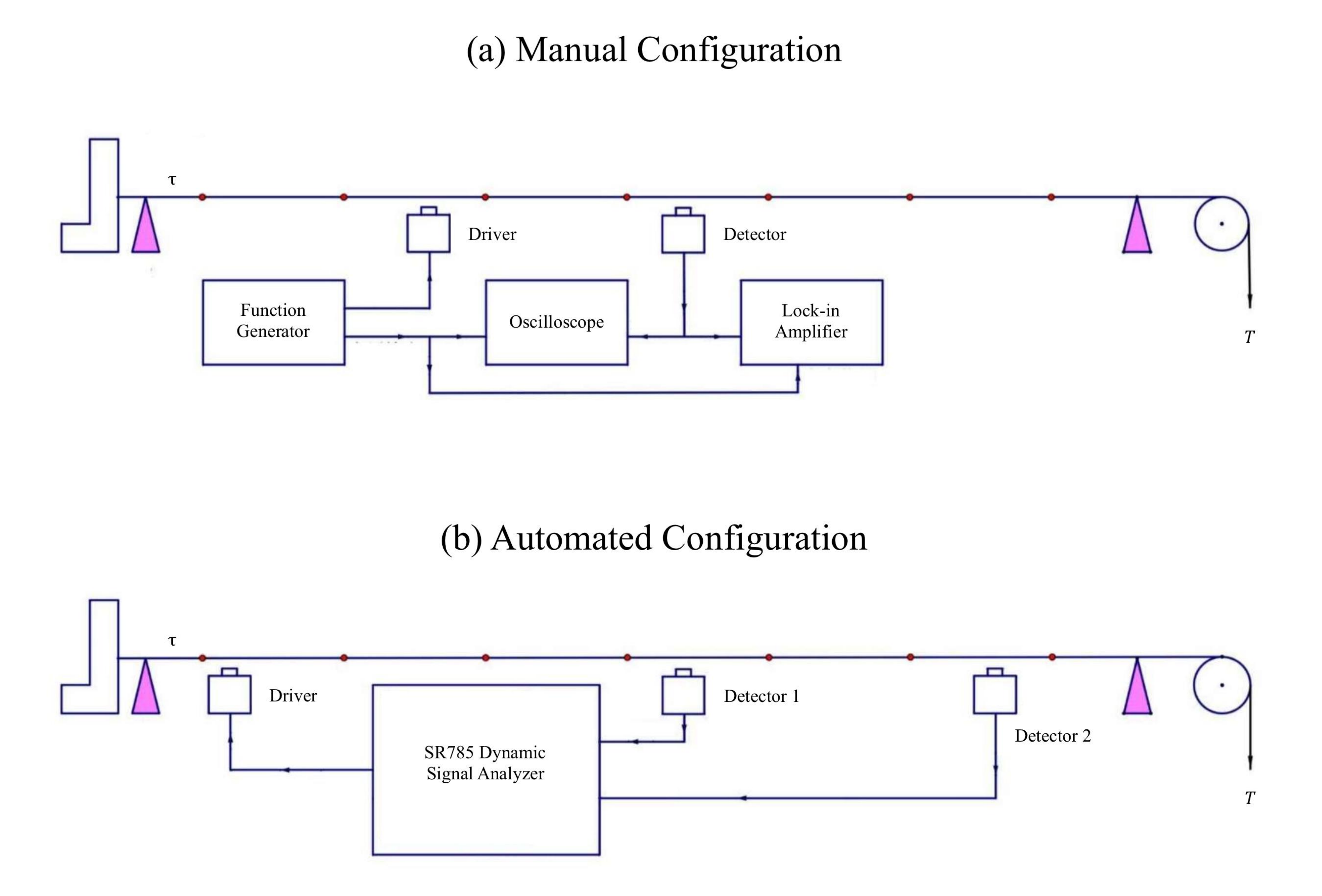}
\caption{Schematic of the experimental setup in (a) manual and (b) automated configurations.
Both share the same mechanical components: beads threaded on a tensioned string, sliding clamps for boundary conditions, and PASCO WA-9613 driver/detector coils.
The manual configuration employs a function generator and lock-in amplifier for detailed mode mapping, while the automated configuration uses an SR785 dynamic signal analyzer for frequency sweeps.}
\label{fig:setup_schematic}
\end{figure}

Our experimental setup follows the design detailed in Ref.~\cite{LiLiu2024}, comprising both manual and automated measurement configurations as illustrated in Fig.~\ref{fig:setup_schematic}.
The core components are identical in both configurations: small beads periodically distributed along a string with linear mass density $\rho_0 = \SI{0.549}{g/m}$ to form a one-dimensional bead-on-string chain.
The string runs above an optical guiding rail, passes over a fixed pulley at one end, and is tensioned by a hanging weight of mass $M$ providing constant tension $T = Mg$.
In our experiment, $T$ is set to $\SI{8.963374}{N}$.
Two sliding clamps at the string ends enforce Dirichlet boundary conditions $U(0)=U(L)=0$ by restricting transverse displacement while leaving the tension unchanged.

Vibration excitation and detection are performed using a PASCO WA-9613 mechanical driver and detector pair, which slide along the optical rail for precise positioning.
In the manual configuration, shown in Fig.~\ref{fig:setup_schematic}(a), a function generator drives the mechanical driver at a single selected eigenfrequency.
A lock-in amplifier then measures the amplitude and phase of the detector signal as the detector is scanned along the string.
This arrangement is optimized for detailed spatial mapping of the mode shape, or eigenfunction, at a specific resonance.
In the automated configuration, shown in Fig.~\ref{fig:setup_schematic}(b), a Stanford Research Systems SR785 dynamic signal analyzer operates in sine-sweep mode to acquire the complete frequency-response spectrum.
This method rapidly identifies eigenfrequencies by scanning a prescribed frequency range and recording the amplitude response at each frequency point.

However, it is important to note that results presented in this paper are obtained from numerical calculations.
The primary reason is that systematic exploration of parameter space requires fine modulation of bead masses and spacings, a level of precision that approaches the limits of our current instrumentation.
To ensure that our numerical codes produce physically meaningful results, we have conducted representative experiments under selected conditions, including uniform and dimerized bead patterns.
The qualitative agreement between measured and calculated mode patterns, as well as the consistency in eigenfrequency ordering, confirms that our computational approach correctly captures the essential physics of the system.

\section{Uniform Periodic Mass Chain\label{sec:uniform}}
\begin{figure}
\includegraphics[width=\linewidth]{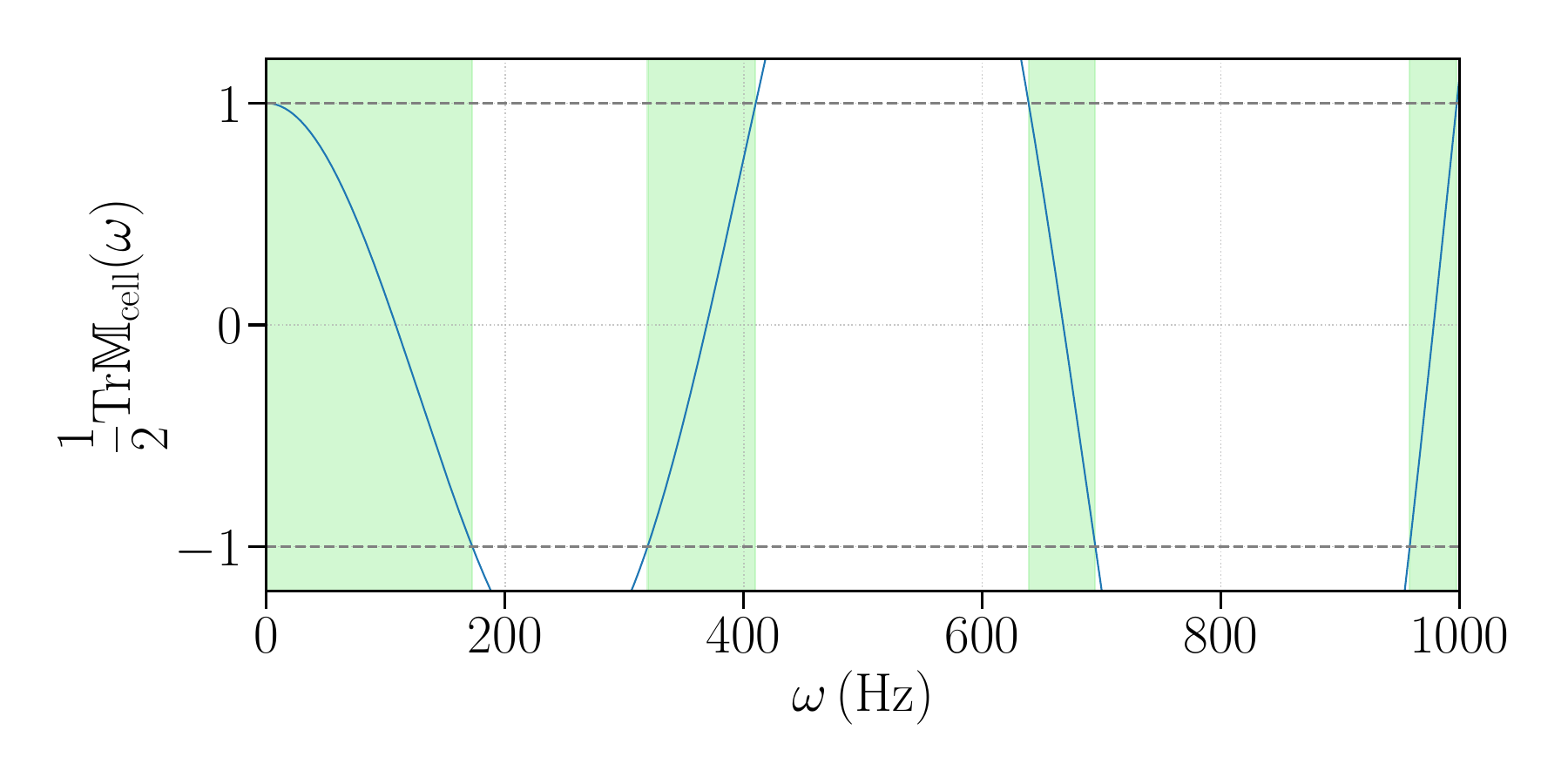}
\caption{Kronig-Penney band diagram for the uniform chain with $a=\SI{20}{cm}$, $T=\SI{8.963374}{N}$, $\rho_0=\SI{0.549}{g/m}$, and $m=\SI{0.114}{g}$.
Shaded regions indicate allowed frequency bands.}
\label{fig:kronig_penney_example}
\end{figure}

We begin with the simplest system: a string of length $L=\SI{120}{cm}$ under tension $T=\SI{8.963374}{N}$ with linear mass density $\rho_0=\SI{0.549}{g/m}$, bearing six identical beads of mass $m=\SI{0.114}{g}$ equally spaced with interval $a=\SI{20}{cm}$.
Fig.~\ref{fig:kronig_penney_example} shows the band structure computed from the transfer-matrix formulation for this uniform periodic chain.

\begin{figure*}
\includegraphics[width=\linewidth]{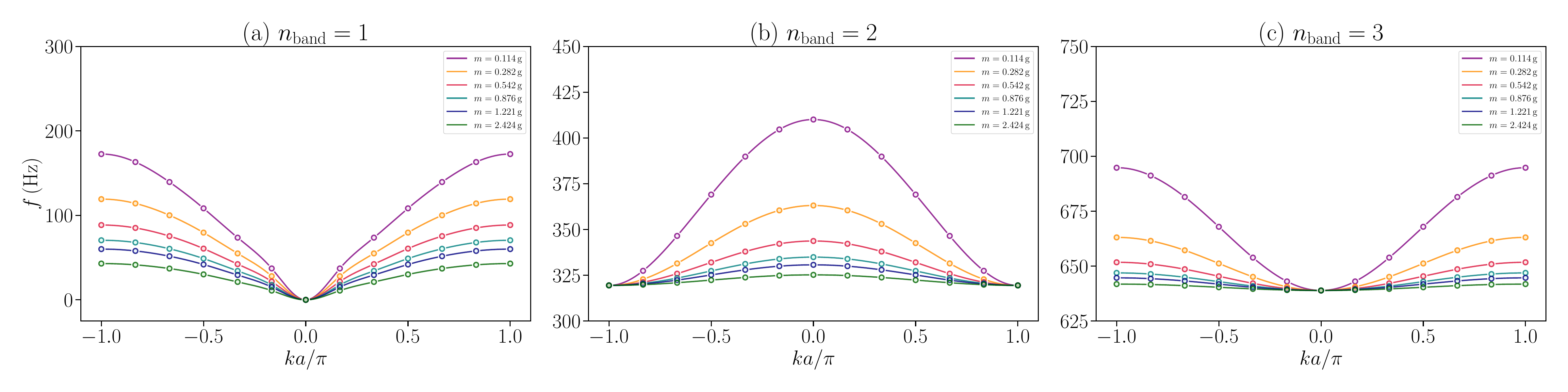}
\caption{Evolution of the first three bands with bead mass $m$.
(a) First band, (b) second band, and (c) third band.
In each panel, the lower edge of the band remains fixed, while the upper edge shifts downward as $m$ increases.
This illustrates that modes near the bottom of a band are insensitive to bead mass, whereas those near the top are strongly affected.}
\label{fig:band_compare}
\end{figure*}

As the bead mass $m$ increases, the higher-frequency parts of each band are progressively suppressed: band tops shift downward while the band bottoms remain strictly fixed, resulting in narrower bands and wider gaps.
Fig.~\ref{fig:band_compare} illustrates this trend for the first three bands.

The insensitivity of band-bottom modes can be understood from a perturbative viewpoint.
When a mode has nodes exactly at the bead positions, its eigenfrequency is unaffected to first order by varying the bead mass $m$.
For illustrative clarity, we consider two simple termination offsets: $\tau=0$ and $\tau=a/2$.

\begin{figure}
\includegraphics[width=\linewidth]{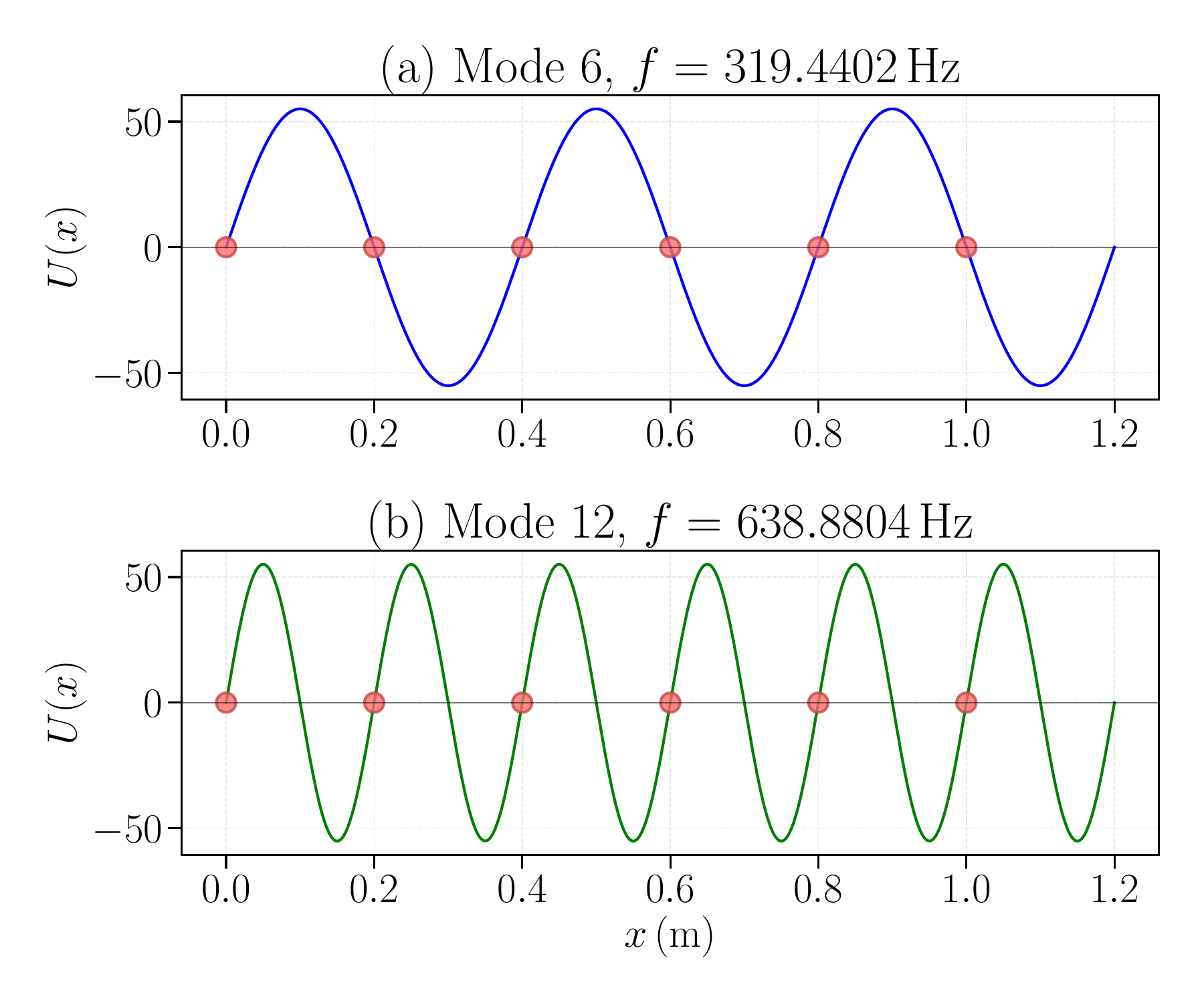}
\caption{Band-bottom modes for termination offset $\tau=0$.
Circles mark the bead positions.
(a) Mode at the bottom of the second band; (b) mode at the bottom of the third band.
In both cases the beads lie exactly at nodes of the displacement, so the eigenfrequencies are insensitive to changes in bead mass.}
\label{fig:band_bottom_modes}
\end{figure}

Fig.~\ref{fig:band_bottom_modes} shows band-bottom modes obtained with $\tau=0$.
In panel (a), corresponding to the bottom of the second band, and in panel (b), corresponding to the bottom of the third band, the beads coincide precisely with nodal points of the displacement profile. Consequently, their eigenfrequencies remain unchanged as the bead mass $m$ varies. 

\begin{figure}
\includegraphics[width=\linewidth]{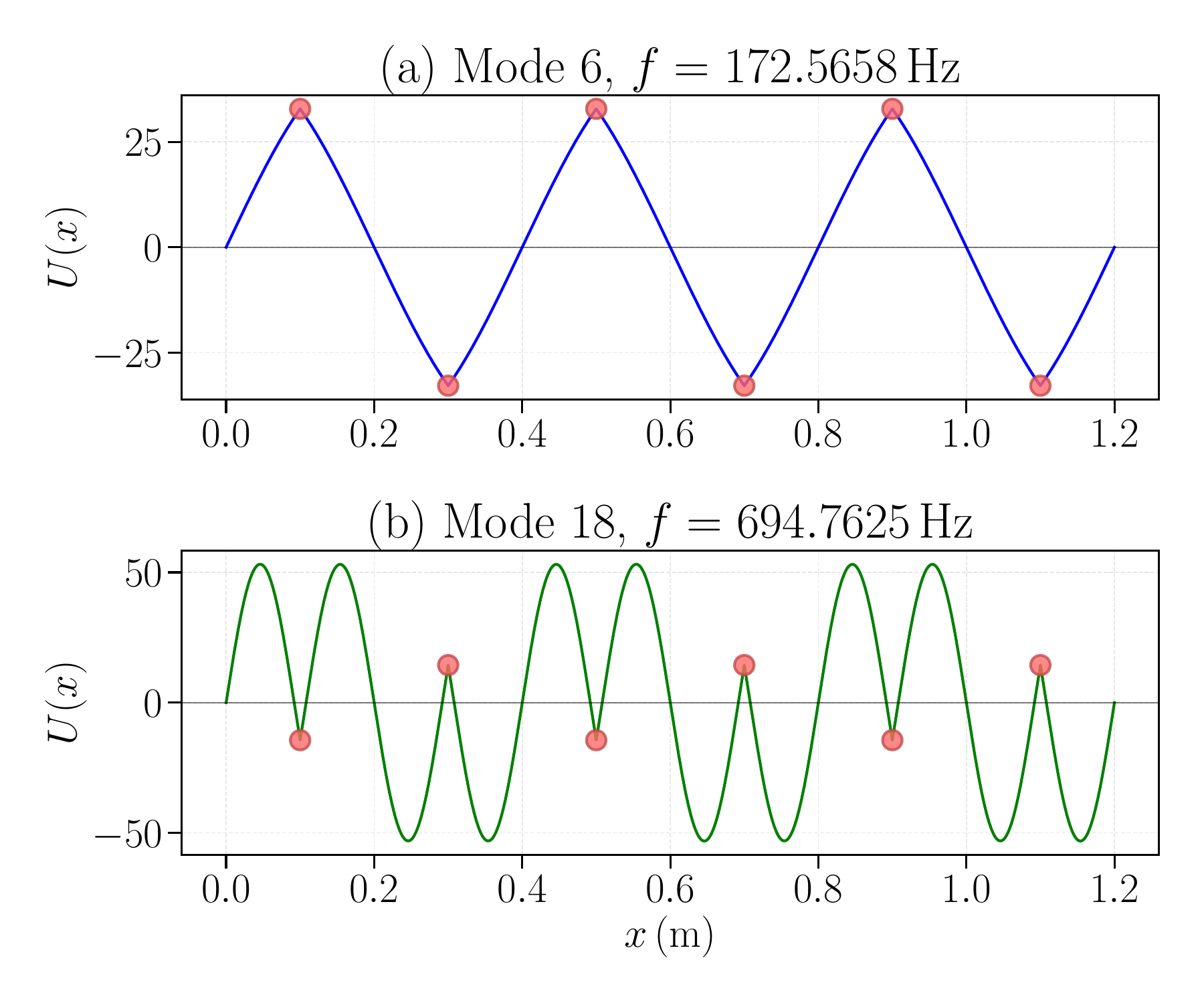}
\caption{Band-top modes for termination offset $\tau=a/2$.
Circles mark the bead positions.
(a) Mode at the top of the first band; (b) mode at the top of the third band.
Here the beads are located at points of nonzero amplitude, causing the eigenfrequencies to shift with bead mass.}
\label{fig:band_top_modes}
\end{figure}

In contrast, Fig.~\ref{fig:band_top_modes} displays band-top modes obtained with $\tau=a/2$.
Panel (a) shows the mode at the top of the first band, and panel (b) shows the mode at the top of the third band.
Here the beads lie at points of finite amplitude, so the frequencies shift downward when $m$ increases.

\begin{figure}
\includegraphics[width=\linewidth]{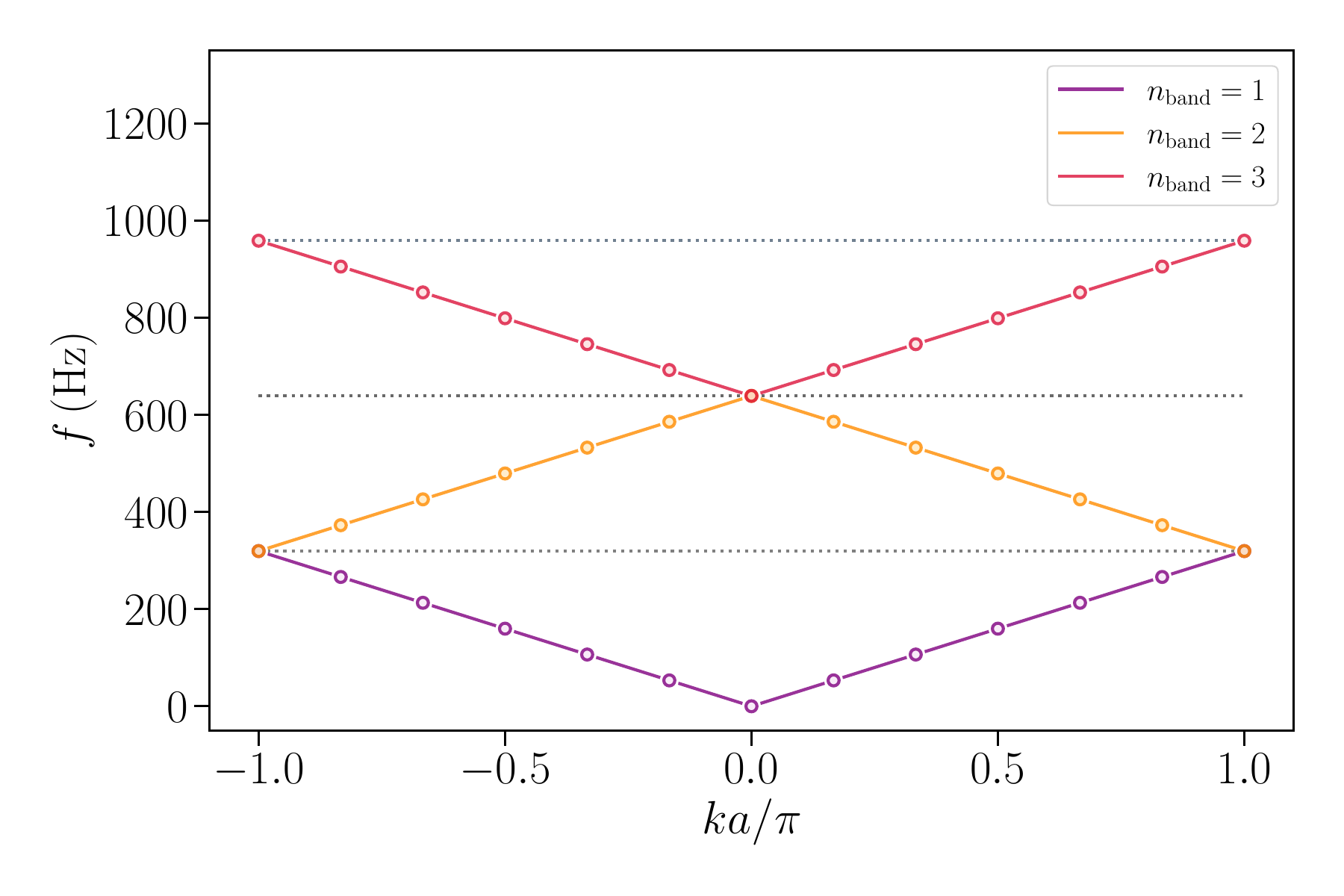}
\caption{Band structure for a uniform string with no beads, corresponding to $m=0$, under the formal period $a=\SI{20}{cm}$.
The curves show the first three bands that would follow from imposing a formal Bloch periodicity.
All gaps vanish, confirming that the gaps observed for nonzero $m$ originate from periodic mass modulation.}
\label{fig:band_no_bead}
\end{figure}

Setting $m=0$ while retaining the formal period $a$ eliminates the periodic mass modulation, causing all band gaps to close as shown in Fig.~\ref{fig:band_no_bead}.
This confirms that the gaps arise from the periodic variation of linear mass density rather than from the mere existence of boundaries.

In solid-state physics, gap closings are a necessary condition for topological transitions, although not every gap closing signals a change of topology.
Some gap closings result from accidental band crossings.
Here the point $m=0$ corresponds to a trivial uniform string with no band structure.
For any nonzero $m$, the periodic mass modulation opens gaps that may host topological edge modes.
In the following sections we test this interpretation by examining whether the midgap states that appear in finite chains for nonzero $m$ satisfy the operational criteria for topological edge states.

\section{Dimerized Periodic Mass Chain and Further Band Modulations\label{sec:dimer_and_further}}
Having examined the uniform periodic chain, we now introduce dimerization: a periodic alternation of two distinct bead masses $m_{\mathrm{A}}$ and $m_{\mathrm{B}}$ within the unit cell.
This doubles the real-space period, halves the size of the first Brillouin zone, and folds the original bands, opening additional gaps in the spectrum.
To realize such a dimerized configuration, we keep all geometric parameters as in the uniform chain but assign alternating masses: odd-indexed beads have $m_{\mathrm{A}}=\SI{0.114}{g}$ and even-indexed beads have $m_{\mathrm{B}}=\SI{0.282}{g}$.
The spatial period therefore increases from $a=\SI{20}{cm}$ to $a=\SI{40}{cm}$.

\begin{figure}
\includegraphics[width=\linewidth]{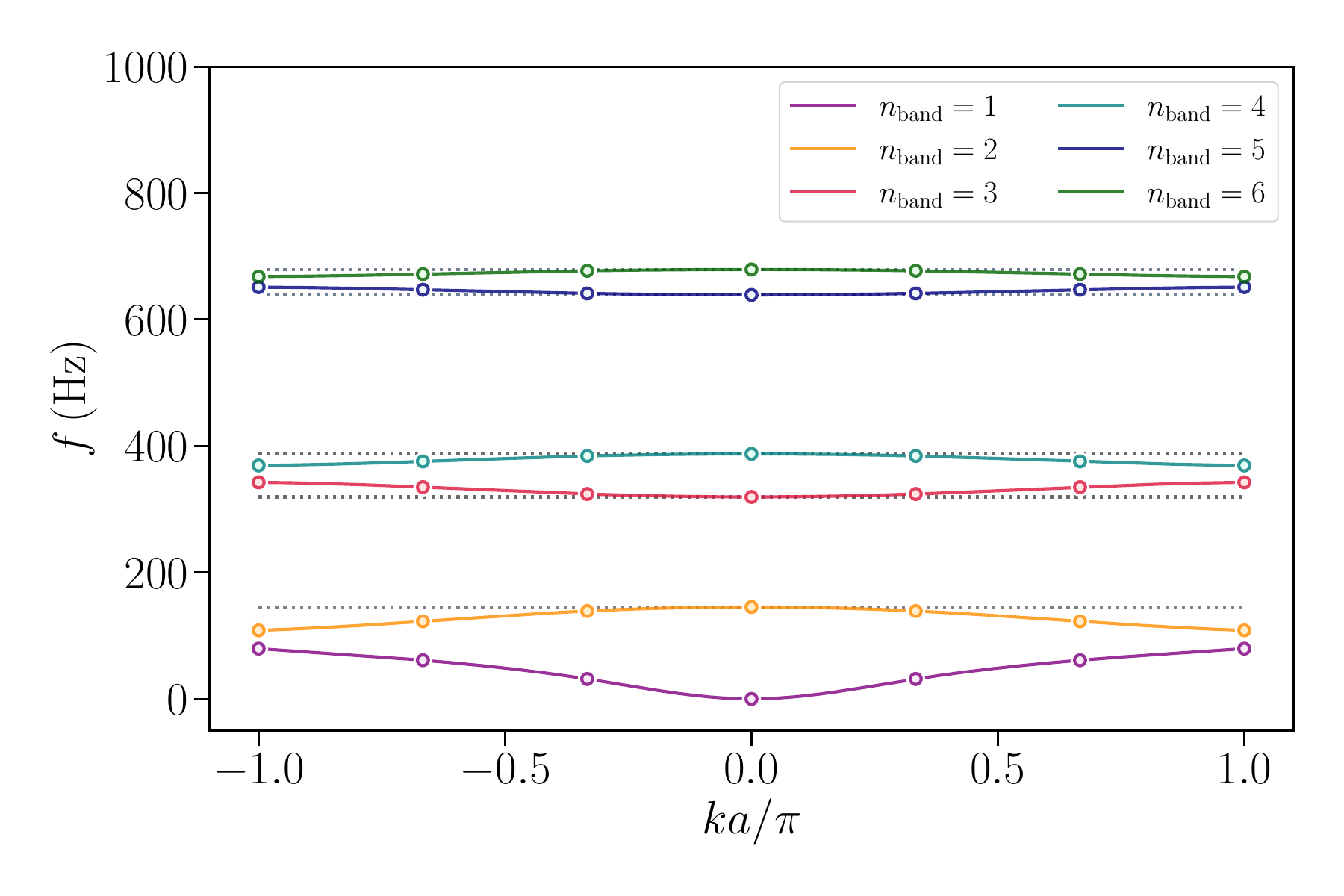}
\caption{Band structure of a dimerized chain with alternating masses $m_{\mathrm{A}}$ and $m_{\mathrm{B}}$ for the first three bands.
Each original band splits into two subbands, creating a new gap inside each original band.}
\label{fig:dimerized_band1}
\end{figure}

Fig.~\ref{fig:dimerized_band1} displays the resulting band structure for the first three bands.
Each of these original bands splits into two subbands, opening a new gap inside each original band.
Consequently, the portion of the spectrum shown now contains six bands separated by five gaps.

\begin{figure}
\includegraphics[width=\linewidth]{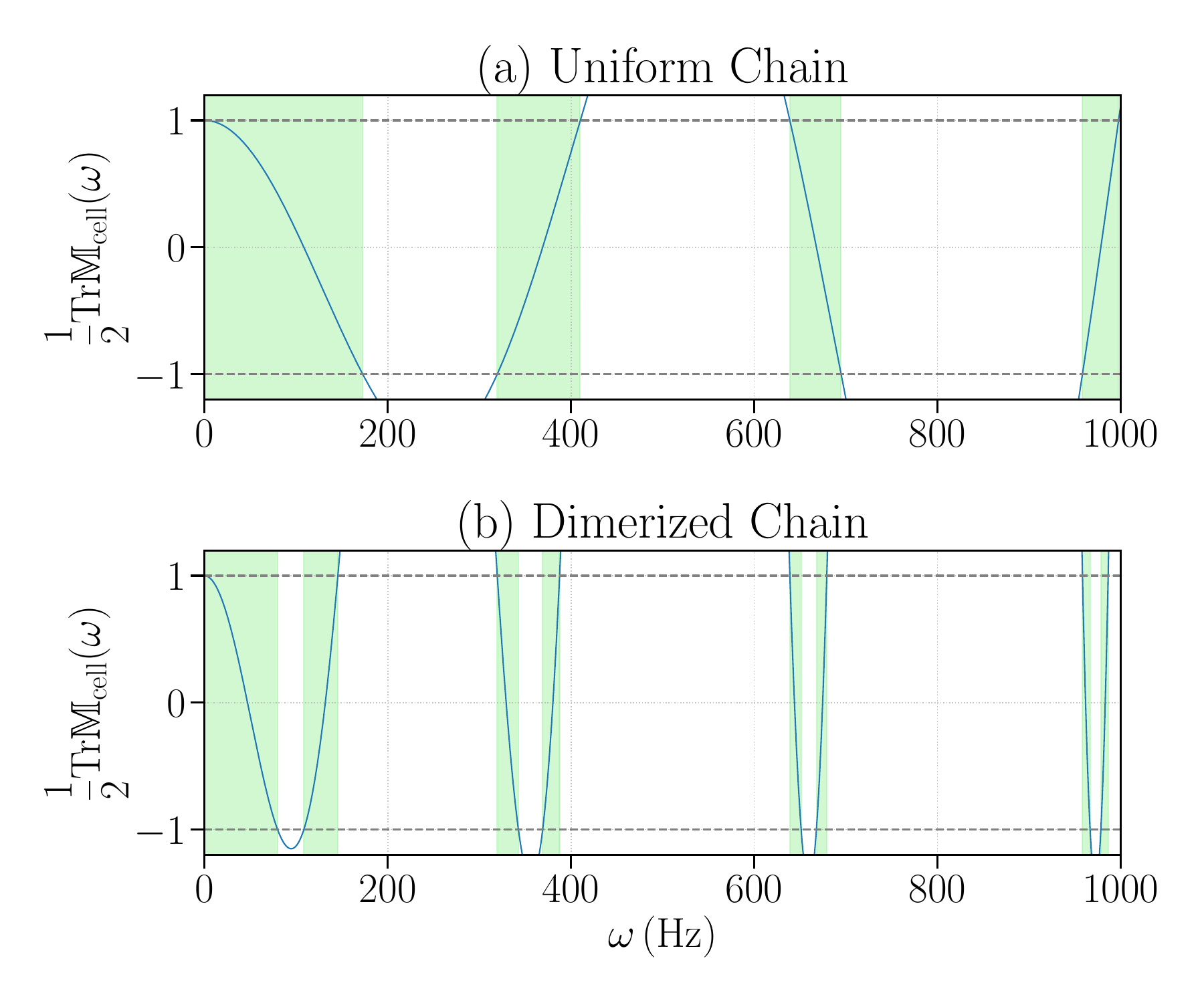}
\caption{Kronig-Penney-type band diagrams computed by the transfer-matrix method.
(a) Uniform chain with $m=\SI{0.114}{g}$, period $a=\SI{20}{cm}$.
(b) Dimerized chain with alternating masses $m_{\mathrm{A}}=\SI{0.114}{g}$ and $m_{\mathrm{B}}=\SI{0.282}{g}$, period $a=\SI{40}{cm}$; in each cell $m_{\mathrm{A}}$ is at $x_{\mathrm{A}}=\SI{0}{cm}$ and $m_{\mathrm{B}}$ at $x_{\mathrm{B}}=\SI{20}{cm}$.
Shaded regions indicate allowed frequency bands.}
\label{fig:dimerized_band_kp}
\end{figure}

A direct comparison between the uniform and dimerized configurations is presented in Fig.~\ref{fig:dimerized_band_kp}, obtained from the transfer-matrix formulation.
Panel (a) shows the uniform chain with $m=\SI{0.114}{g}$ and period $a=\SI{20}{cm}$.
Panel (b) shows the dimerized chain with alternating masses $m_{\mathrm{A}}$ and $m_{\mathrm{B}}$ and period $a=\SI{40}{cm}$, where in each unit cell $m_{\mathrm{A}}$ is placed at $x_{\mathrm{A}}=\SI{0}{cm}$ and $m_{\mathrm{B}}$ at $x_{\mathrm{B}}=\SI{20}{cm}$.
The dimerization clearly splits each original allowed band and introduces additional gaps.

To probe how the dimerized band structure depends on intra-cell geometry, we now take the unit cell to span $[0,40)\,\mathrm{cm}$ and keep the heavy bead $m_{\mathrm{B}}$ at $x=\SI{20}{cm}$, while varying the position $x_{\mathrm{A}}$ of the light bead $m_{\mathrm{A}}$ within the cell.

\begin{figure}
\includegraphics[width=\linewidth]{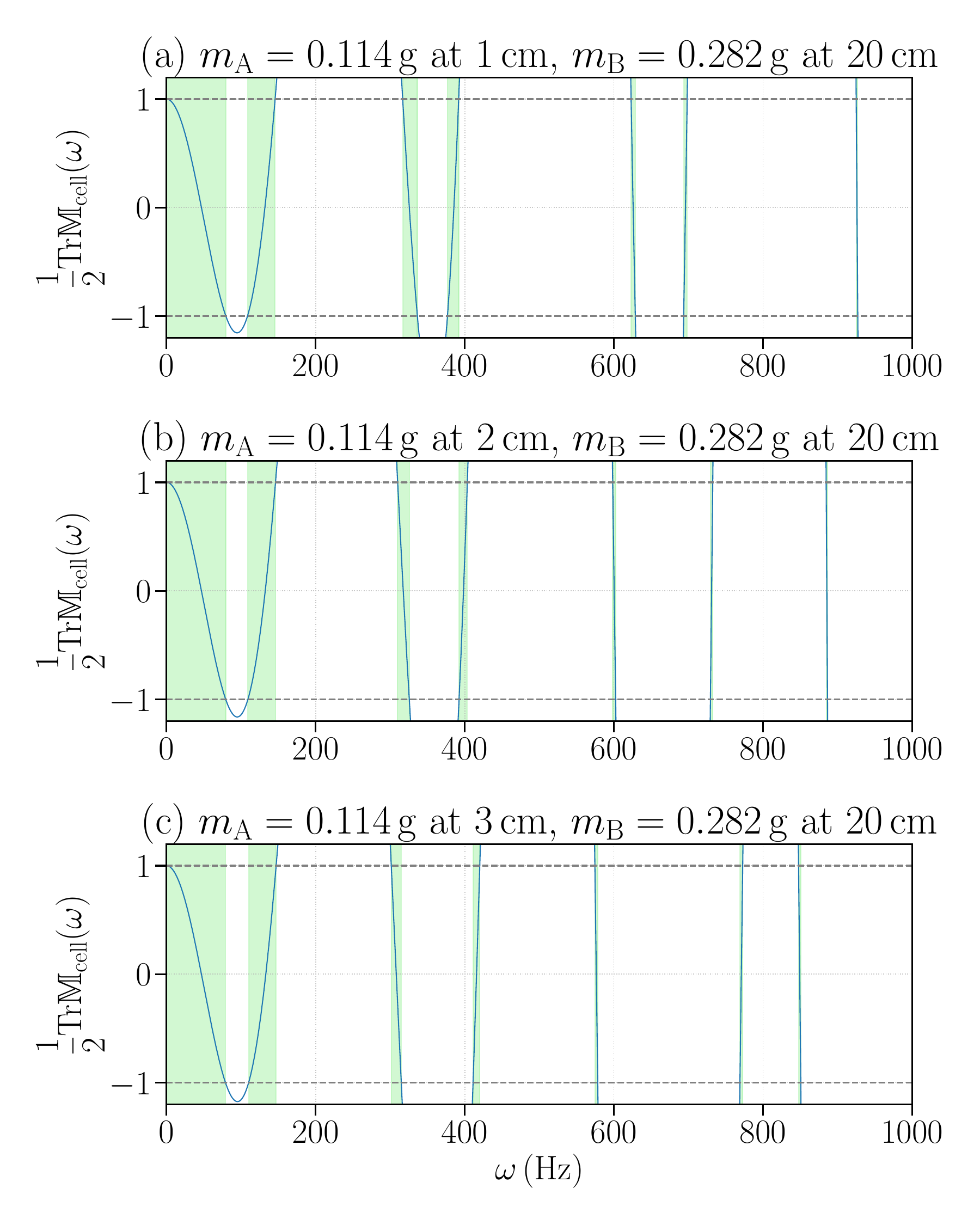}
\caption{Kronig-Penney-type band diagrams computed by the transfer-matrix method for the dimerized chain with $m_{\mathrm{A}}=\SI{0.114}{g}$ and $m_{\mathrm{B}}=\SI{0.282}{g}$, period $a=\SI{40}{cm}$, $x_{\mathrm{B}}=\SI{20}{cm}$ fixed.
(a) $x_{\mathrm{A}}=\SI{1}{cm}$, (b) $x_{\mathrm{A}}=\SI{2}{cm}$, (c) $x_{\mathrm{A}}=\SI{3}{cm}$.
Shaded regions indicate allowed frequency bands.}
\label{change_position_123}
\end{figure}

Fig.~\ref{change_position_123} shows the resulting Kronig-Penney band diagrams for three closely spaced positions: $x_{\mathrm{A}}=1$, $2$, and $3\,\mathrm{cm}$.
As the light bead moves away from the left boundary, the intra-cell spacing decreases and the coupling strengthens, visibly widening the extra-gaps inside each original band.

\begin{figure}
\includegraphics[width=\linewidth]{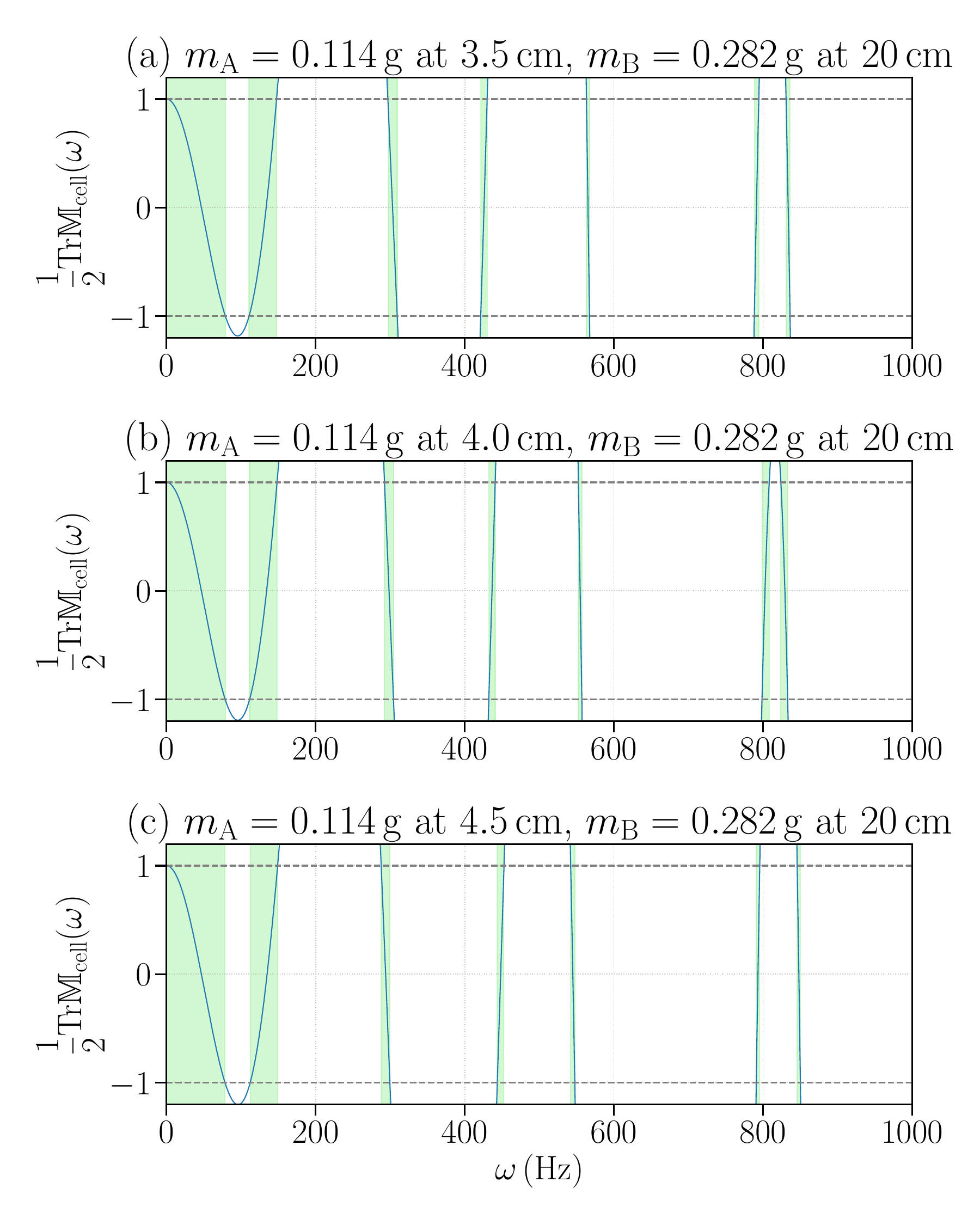}
\caption{Kronig-Penney-type band diagrams computed by the transfer-matrix method for the dimerized chain with $m_{\mathrm{A}}=\SI{0.114}{g}$ and $m_{\mathrm{B}}=\SI{0.282}{g}$, period $a=\SI{40}{cm}$, $x_{\mathrm{B}}=\SI{20}{cm}$ fixed.
(a) $x_{\mathrm{A}}=\SI{3.5}{cm}$, (b) $x_{\mathrm{A}}=\SI{4.0}{cm}$, (c) $x_{\mathrm{A}}=\SI{4.5}{cm}$.
At $x_{\mathrm{A}}\simeq\SI{4}{cm}$ a gap closing occurs between the second subband of the original third band and the first subband of the original fourth subband.
Shaded regions indicate allowed frequency bands.}
\label{change_position_345}
\end{figure}

This widening causes the subbands to shift relative to each other; they can approach and eventually touch when a gap closes.
Fig.~\ref{change_position_345} provides a detailed scan around $x_{\mathrm{A}}\simeq\SI{4}{cm}$, where such a gap closing occurs between the second subband of the original third band and the first subband of the original fourth subband.
These band touchings are often accidental: they result simply from the changing overlap of subbands as the intra-cell geometry varies, without an exchange of band character, and therefore do not necessarily imply a topological transition.

\begin{figure*}
\includegraphics[width=\linewidth]{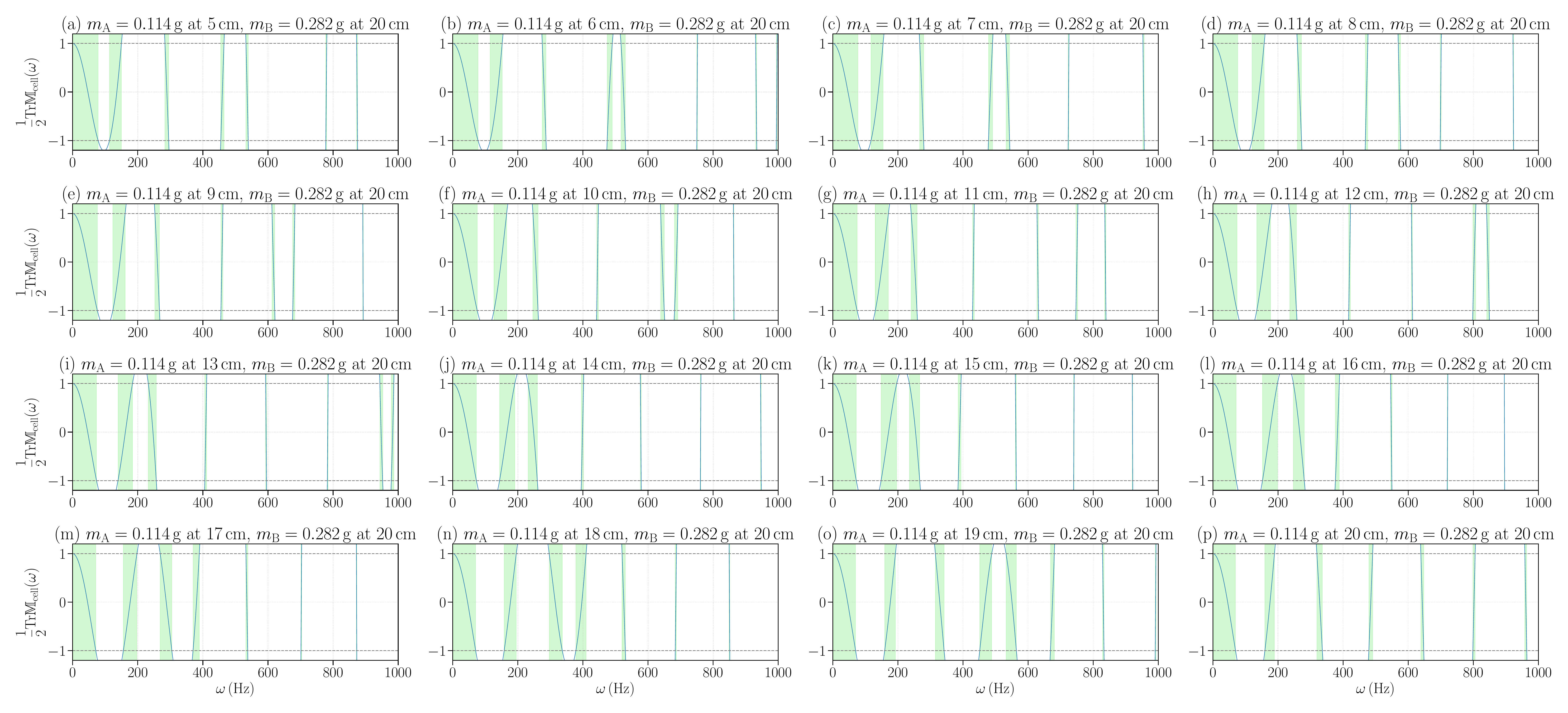}
\caption{Comprehensive Kronig-Penney-type band diagrams for the dimerized chain with $m_{\mathrm{A}}=\SI{0.114}{g}$ and $m_{\mathrm{B}}=\SI{0.282}{g}$, period $a=\SI{40}{cm}$, $x_{\mathrm{B}}=\SI{20}{cm}$ fixed.
Panels (a)-(p) correspond to $x_{\mathrm{A}}=5,6,\dots,20\,\mathrm{cm}$ in $1\,\mathrm{cm}$ steps.
Shaded regions indicate allowed frequency bands.}
\label{change_position_more}
\end{figure*}

The full evolution from $x_{\mathrm{A}}=\SI{5}{cm}$ to $\SI{20}{cm}$ is displayed in Fig.~\ref{change_position_more}.
Sixteen panels show how the subbands systematically shift and reconnect as the light bead approaches the heavy bead at the cell center.
When the two beads coincide at $x_{\mathrm{A}}=\SI{20}{cm}$, the system reverts to a uniform chain with period $a=\SI{40}{cm}$ and effective mass $m=m_{\mathrm{A}}+m_{\mathrm{B}}$.

The dimerized bead-on-string system shares qualitative features with the SSH model, where alternating hoppings $t_1$ and $t_2$ control the topological phase under the condition of equal onsite energies.
In the SSH model the topological transition occurs at $t_1 = t_2$ while keeping the onsite terms identical. 

A direct mapping to our mechanical system is subtle because both mass and spacing influence the effective hopping.
Moreover, varying the mass also changes the onsite energy, making it difficult to isolate the hopping modulation as in the SSH model.
To circumvent this complication, we focus on the case where only the spacing is varied while the beads remain identical.
Here the onsite energies stay equal by symmetry, while differences in intra-cell and inter-cell spacings generate distinct effective hoppings $t_1$ and $t_2$.
This configuration mimics the SSH condition of equal onsite energies with tunable hoppings, and a gap closing is expected when the two spacings become equal.

\begin{figure*}
\includegraphics[width=\linewidth]{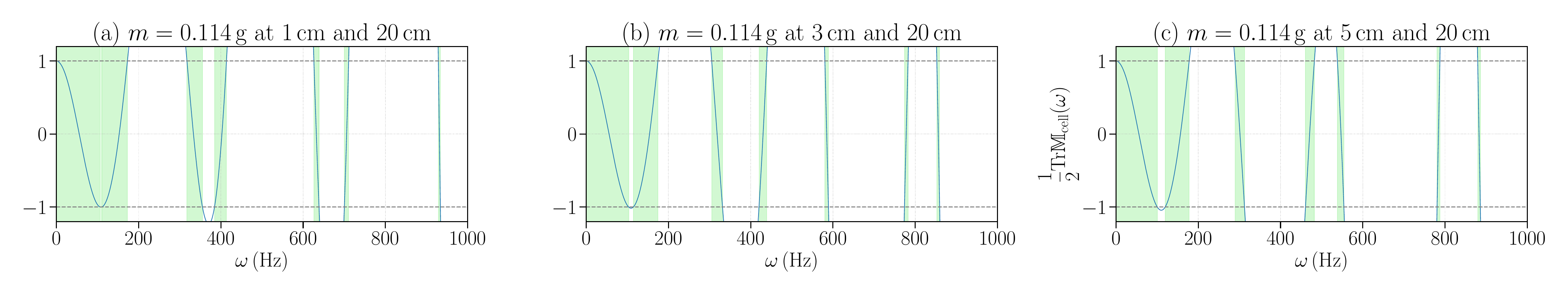}
\caption{Kronig-Penney-type band diagrams computed by the transfer-matrix method for a dimerized unit cell with two identical beads $m=\SI{0.114}{g}$ placed at $\SI{20}{cm}$ and at (a) $\SI{1}{cm}$, (b) $\SI{3}{cm}$, (c) $\SI{5}{cm}$.
Varying the relative spacing opens extra-gaps inside each original band, demonstrating that geometric dimerization alone can produce band gaps.
Shaded regions indicate allowed frequency bands.}
\label{fig:same_bead_distance_19_17_15}
\end{figure*}

To demonstrate this geometric route, we again take a unit cell spanning $[0,40)\,\mathrm{cm}$ and place one bead of mass $m=\SI{0.114}{g}$ at $x=\SI{20}{cm}$.
The other bead, with the same mass, is placed at $x=1$, $3$, or $5\,\mathrm{cm}$.
Fig.~\ref{fig:same_bead_distance_19_17_15} shows the resulting Kronig-Penney band diagrams for these three configurations.
Even without a mass contrast, the asymmetry in intra-cell spacing splits each original band and opens extra-gaps.
This confirms that a purely geometric dimerization can generate band gaps, establishing a mechanical analog of the SSH condition where onsite energies are equal but hoppings differ.

\section{Edge-State Modes and Determination of Topological Transitions\label{sec:topo_transit}}
Building on the infinite-cell band analysis introduced above, we now examine finite-chain spectra to identify potential topological edge states according to the diagnostic criteria.
For each candidate mode we test: (1) adiabatic continuity with respect to the termination offset $\tau$, and (2) robustness to small, localized perturbations at the boundary.

\subsection{Uniform periodic chain: termination dependence and edge modes\label{subsec:uniform_edge}}
\begin{figure}
\includegraphics[width=\linewidth]{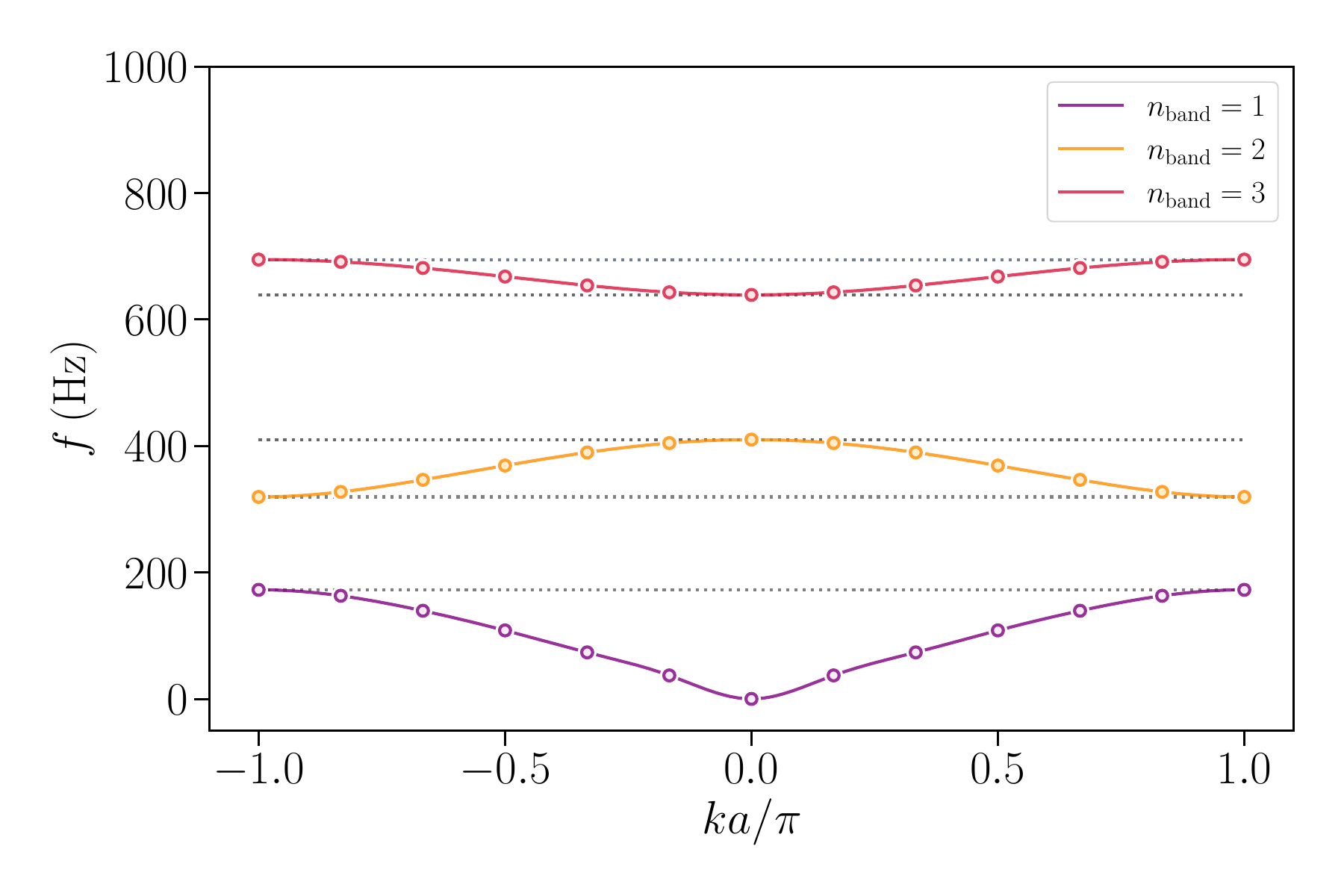}
\caption{Band structure of the uniform chain with $a=\SI{20}{cm}$, $T=\SI{8.963374}{N}$, $\rho_0=\SI{0.549}{g/m}$, and $m=\SI{0.114}{g}$.}
\label{band_1}
\end{figure}

We begin with the simplest configuration: a finite string of length $L=\SI{120}{cm}$ under tension $T=\SI{8.963374}{N}$ with linear mass density $\rho_0=\SI{0.549}{g/m}$, bearing six identical beads of mass $m=\SI{0.114}{g}$ equally spaced with interval $a=\SI{20}{cm}$.
Fig.~\ref{band_1} shows the band structure for this uniform chain.

The band structure in Fig.~\ref{band_1} reveals three bulk gaps that can potentially support edge states.
In a finite chain, whether such states actually appear and at what frequencies depends sensitively on how the periodic pattern meets the boundaries.
To quantify this dependence, we vary the termination offset $\tau$ and track the resulting midgap eigenfrequencies.

\begin{figure}
\includegraphics[width=\linewidth]{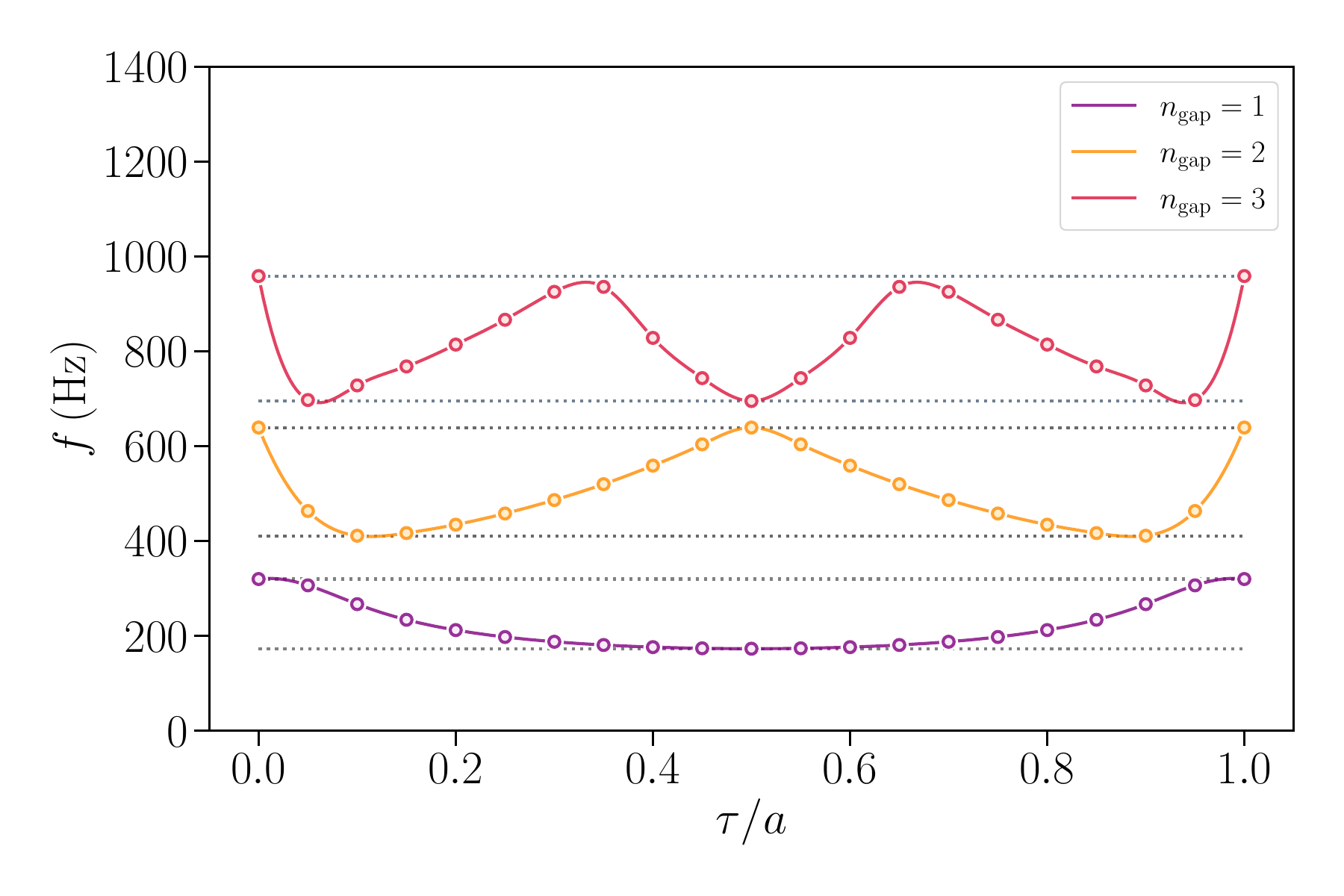}
\caption{Edge-state eigenfrequencies in the uniform chain as a function of the termination offset $\tau$.
Symmetry points $\tau=0$ and $\tau=a/2$ suppress edge states, whereas asymmetric terminations allow localized modes to emerge.}
\label{fig:uniform_edge_overview}
\end{figure}

Fig.~\ref{fig:uniform_edge_overview} summarizes the dependence of midgap eigenfrequencies on the termination offset $\tau$.
At symmetric terminations $\tau=0$ and $\tau=a/2$, the gaps do not support edge modes, while for asymmetric terminations localized gap modes appear and trace continuous curves as $\tau$ varies.

The parameter scans in Fig.~\ref{fig:uniform_edge_overview} show that the candidate midgap modes located in the first, second and third gaps, labeled as modes 6, 12 and 18, evolve smoothly with $\tau$ and remain spectrally isolated from the bulk bands.
All three candidates therefore satisfy the first diagnostic requirement: they appear within reopened gaps and persist adiabatically as the termination parameter $\tau$ is varied. 

To further determine their topological character, we now test the second diagnostic requirement: robustness to local boundary perturbations.
For each candidate mode, we apply two distinct perturbations: a left-side perturbation, where the leftmost bead is shifted $\SI{5}{cm}$ to the right, and a right-side perturbation, where the rightmost bead is shifted $\SI{5}{cm}$ to the left.
These displacements are small compared with the chain length $L=\SI{120}{cm}$ and the inter-bead spacing $a=\SI{20}{cm}$, and they leave the global tension unchanged.
They therefore probe only the local sensitivity of the edge modes to boundary details.
All unperturbed configurations used in these tests have termination offset $\tau=a/4=\SI{5}{cm}$.

\begin{figure*}
\includegraphics[width=\linewidth]{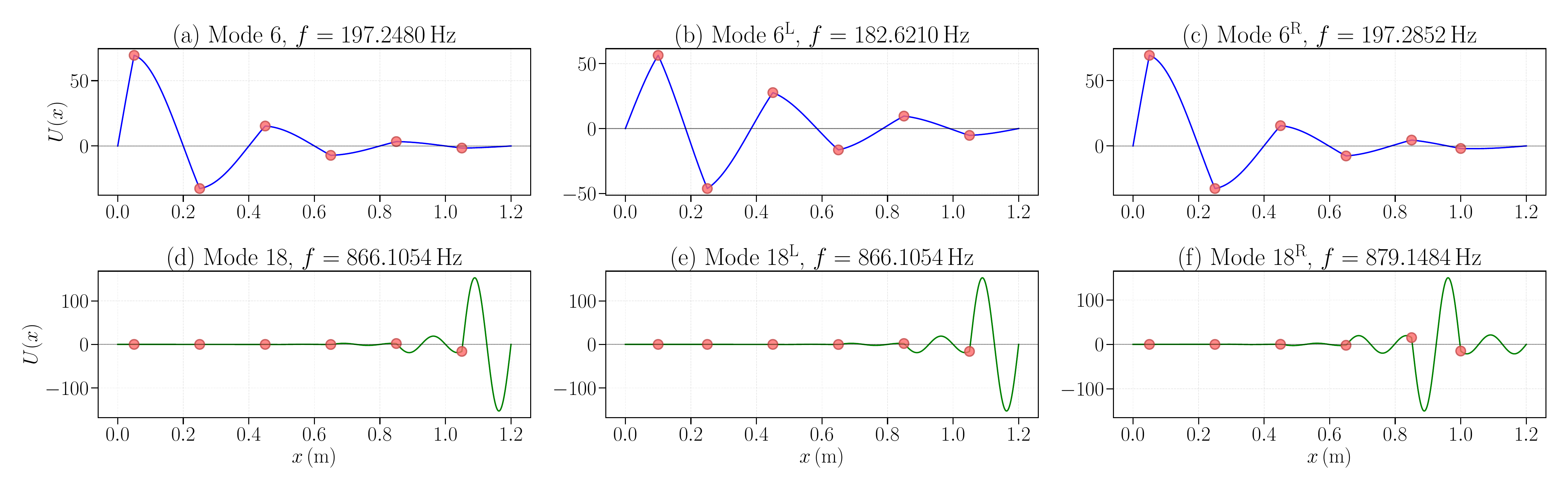}
\caption{Responses of modes 6 and 18 to boundary perturbations for the uniform chain with $\tau=a/4=\SI{5}{cm}$.
(a) Unperturbed mode 6; (b) mode 6 after left-side perturbation; (c) mode 6 after right-side perturbation.
(d) Unperturbed mode 18; (e) mode 18 after left-side perturbation; (f) mode 18 after right-side perturbation.
Mode 6, localized on the left, is robust to the right-side perturbation; mode 18, localized on the right, is robust to the left-side perturbation.
In the plot, the blue solid line shows the mode profile, and red circles denote beads with $m_{\mathrm{A}}=\SI{0.114}{g}$.}
\label{fig:true_topo_edge_modes}
\end{figure*}

Fig.~\ref{fig:true_topo_edge_modes} displays the responses of modes 6 and 18 to the boundary perturbations.
Panel (a) shows the unperturbed mode 6, localized near the left end of the chain, and panel (d) shows the unperturbed mode 18, localized near the right end.
In panels (b) and (e) we apply the left-side perturbation; in panels (c) and (f) we apply the right-side perturbation.
Mode 6, already localized on the left, remains essentially unchanged under the right-side perturbation and is nearly unaffecteds by the left-side perturbation.
Similarly, mode 18, localized on the right, withstands the left-side perturbation and shows only minor adjustment under the right-side perturbation.
Both modes stay firmly inside their respective gaps and retain their exponentially localized profiles.

\begin{figure*}
\includegraphics[width=\linewidth]{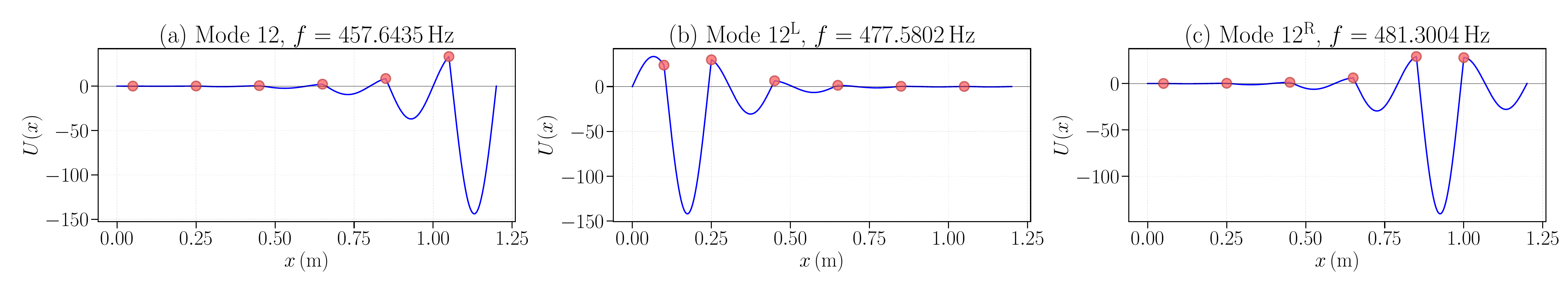}
\caption{Response of mode 12 to boundary perturbations for the uniform chain with $\tau=a/4=\SI{5}{cm}$.
(a) Unperturbed mode 12; (b) after left-side perturbation; (c) after right-side perturbation.
The mode is substantially altered by both perturbations.
In the plot, the blue solid line shows the mode profile, and red circles denote beads with $m_{\mathrm{A}}=\SI{0.114}{g}$.}
\label{fig:fake_topo_edge_modes}
\end{figure*}

Fig.~\ref{fig:fake_topo_edge_modes} shows the corresponding tests for mode 12.
Panel (a) displays the unperturbed mode, which is localized on the right side.
Panel (b) presents the result after the left-side perturbation, and panel (c) after the right-side perturbation.
Unlike modes 6 and 18, the spatial profile of mode 12 is strongly affected by both perturbations.
This sensitivity indicates that mode 12 does not originate from bulk topology but rather from the specific termination at the boundary.

Combining the adiabatic-continuity test with the perturbation tests, we classify the edge modes as follows.
Mode 6 in the first gap and mode 18 in the third gap satisfy both diagnostic requirements: they evolve smoothly with termination offset $\tau$ and remain robust under local boundary modifications.
We therefore identify them as bulk-driven topological edge states.
In contrast, mode 12 in the second gap, although it also evolves adiabatically with $\tau$, fails the robustness test; it is consequently a termination-induced boundary resonance with no topological significance.

\subsection{Dimerized chain: band folding and SSH mapping\label{subsec:dimer_edge}}
We introduce dimerization by placing beads with two different masses $m_{\mathrm{A}}=\SI{0.114}{g}$ and $m_{\mathrm{B}}=\SI{0.282}{g}$ in an alternating pattern.
This creates a two-site unit cell that doubles the period and folds the bands, opening additional gaps as shown in Fig.~\ref{fig:dimerized_band1}.

\begin{figure}
\includegraphics[width=\linewidth]{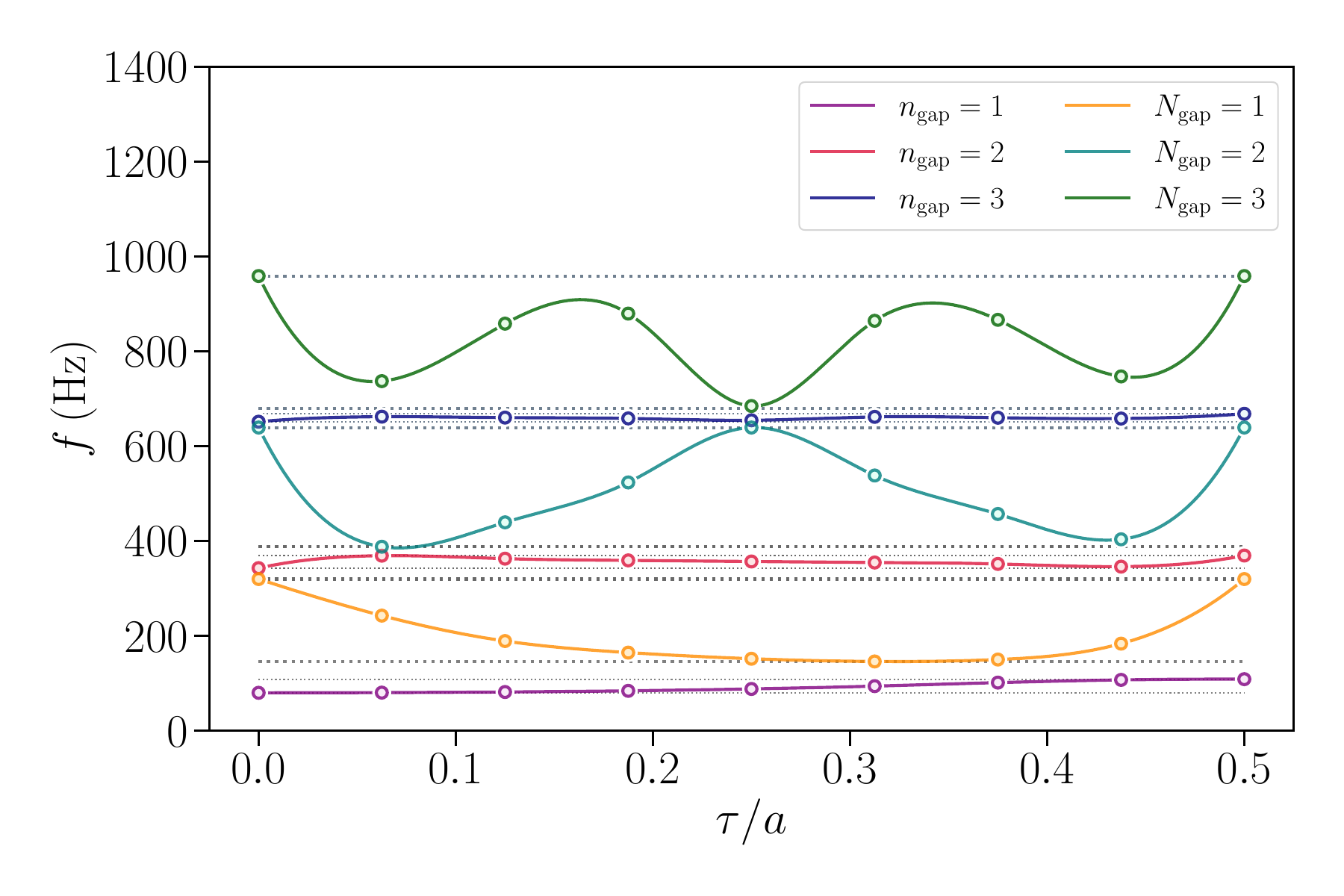}
\caption{Edge-state eigenfrequencies in the dimerized chain as a function of the termination offset $\tau$.
Extra gaps created by dimerization support additional localized modes.
Symmetry points $\tau=0$ and $\tau=a/2$ suppress edge states, whereas asymmetric terminations allow localized modes to emerge.}
\label{fig:dimerized_edge_freq}
\end{figure}

Edge-state eigenfrequencies in the dimerized chain exhibit richer structure than the uniform case.
Fig.~\ref{fig:dimerized_edge_freq} plots the eigenfrequencies of midgap states as a function of the termination offset $\tau$, where the unit cell length is $a=\SI{40}{cm}$.
In addition to the original gaps, the dimerization opens new gaps, each of which supports at most one robust midgap mode for asymmetric terminations.

As in the uniform-chain case, parameter scans for the dimerized unit cell show that the candidate midgap modes evolve smoothly and remain spectrally isolated from the bulk bands, satisfying the first diagnostic requirement. 
To determine which of these candidates are bulk-driven and which are termination-induced, we subject three representative edge states from the extra-gaps to localized boundary displacements, with the termination offset fixed at $\tau=a/8=\SI{5}{cm}$.
The three candidates are mode 3 from the first extra-gap, mode 9 from the second extra-gap, and mode 15 from the third extra-gap.

\begin{figure*}
\includegraphics[width=\linewidth]{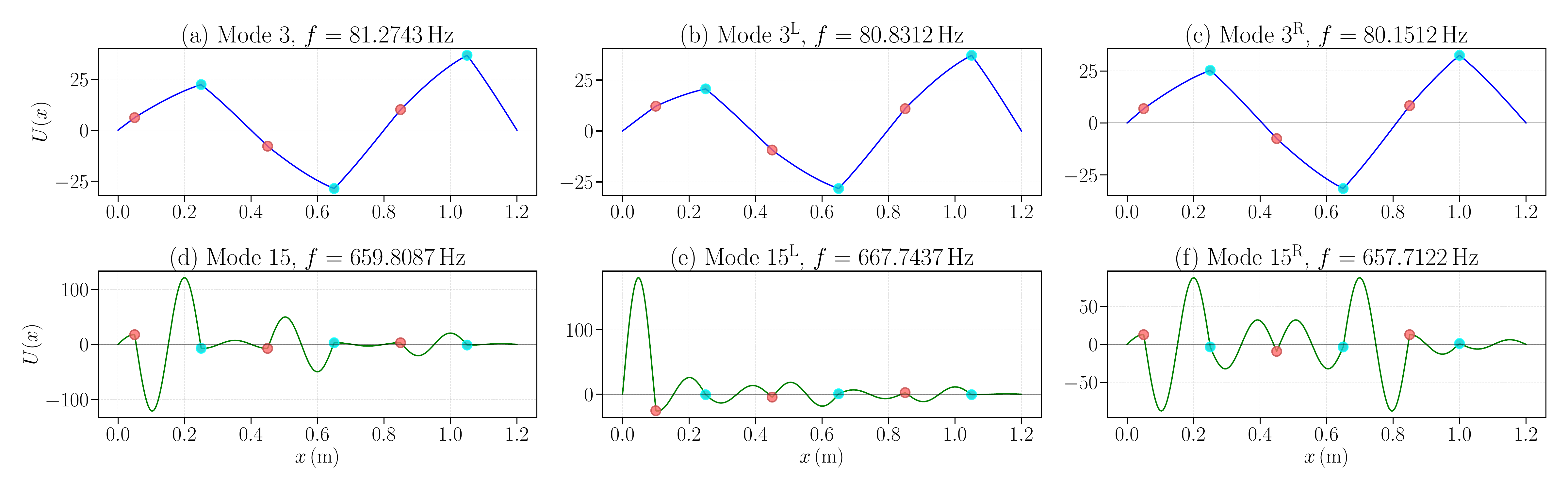}
\caption{Responses of modes 3 and 15 to boundary perturbations for the dimerized chain with $\tau=a/8=\SI{5}{cm}$.
(a) Unperturbed mode 3; (b) mode 3 after left-side perturbation; (c) mode 3 after right-side perturbation.
(d) Unperturbed mode 15; (e) mode 15 after left-side perturbation; (f) mode 15 after right-side perturbation.
Mode 3, localized on the right, is robust to the left-side perturbation; mode 15, localized on the left, is robust to the right-side perturbation.
In the plot, the blue solid line shows the mode profile, red circles denote light beads with $m_{\mathrm{A}}=\SI{0.114}{g}$, and cyan circles denote heavy beads with $m_{\mathrm{B}}=\SI{0.282}{g}$.}
\label{fig:dimer_true_topo_edge_modes}
\end{figure*}

Fig.~\ref{fig:dimer_true_topo_edge_modes} shows the responses of modes 3 and 15 to boundary perturbations.
Panel (a) displays the unperturbed mode 3, which is localized near the right end of the chain.
Panel (d) displays the unperturbed mode 15, which is localized near the left end.
Panels (b) and (e) show the modes after shifting the leftmost bead $\SI{5}{cm}$ inward, while panels (c) and (f) show the modes after shifting the rightmost bead $\SI{5}{cm}$ inward.
Mode 3, localized on the right, remains essentially unchanged under the left-side perturbation and is nearly unaffected by the right-side perturbation.
Similarly, mode 15, localized on the left, is robust to the right-side perturbation as the perturbation cannot disrupt its left-localized profile.
Both modes stay firmly inside their respective gaps and retain exponentially localized profiles.

\begin{figure*}
\includegraphics[width=\linewidth]{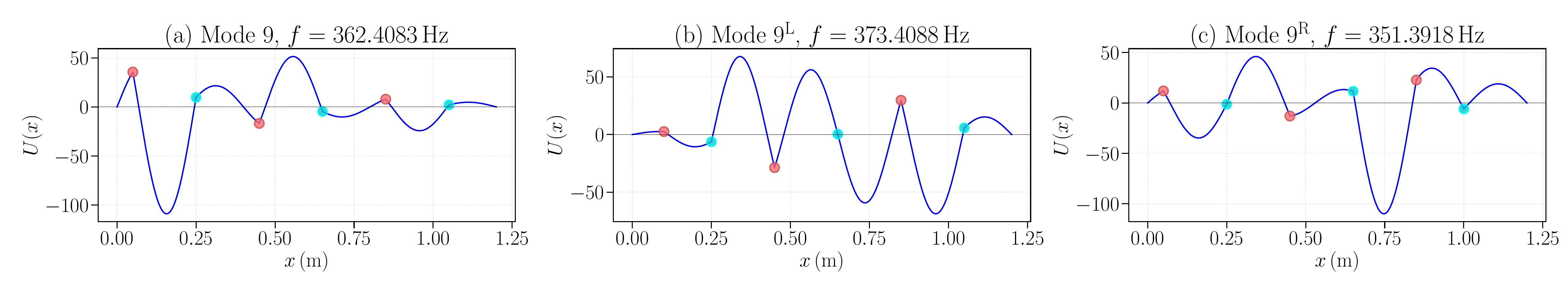}
\caption{Response of mode 9 to boundary perturbations for the dimerized chain with $\tau=a/8=\SI{5}{cm}$.
(a) Unperturbed mode 9; (b) after left-side perturbation; (c) after right-side perturbation.
The mode is substantially altered by both perturbations.
In the plot, the blue solid line shows the mode profile, red circles denote light beads with $m_{\mathrm{A}}=\SI{0.114}{g}$, and cyan circles denote heavy beads with $m_{\mathrm{B}}=\SI{0.282}{g}$.}
\label{fig:dimer_fake_topo_edge_modes}
\end{figure*}

Fig.~\ref{fig:dimer_fake_topo_edge_modes} shows the corresponding tests for mode 9.
Panel (a) displays the unperturbed mode, which is localized on the left side.
Panel (b) presents the result after the left-side perturbation, and panel (c) after the right-side perturbation.
Unlike modes 3 and 15, the spatial profile of mode 9 is strongly affected by both perturbations.
This sensitivity indicates that mode 9 does not originate from bulk topology but rather from the specific termination at the boundary.

Combining the adiabatic-continuity test with the perturbation tests, we classify mode 3 in the first extra-gap and mode 15 in the third extra-gap as bulk-driven topological edge states, as they satisfy both diagnostic requirements.
In contrast, mode 9 in the second extra-gap, although it evolves adiabatically with $\tau$, fails the robustness test; it is consequently a termination-induced boundary resonance with no topological significance.

This pattern is consistent with what we observed in the uniform chain: edge states in the first and third gaps are robust, while those in the second gap are fragile.
In both uniform and dimerized configurations, we find that the first and third gaps support topological edge states, whereas the second gap yields termination-induced modes.
This systematic distinction will be explained in the following section through an SSH-like two-band reduction and the corresponding (1+1)-dimensional Dirac mass-domain picture.

\section{Domain Walls and Dirac Mapping\label{sec:domain_wall}}
The numerical and experimental results presented above motivate a compact continuum-level explanation.
We have observed that in both uniform and dimerized chains, the first and third gaps support robust edge states while the second gap yields fragile termination-induced modes.
This systematic pattern can be understood by noting that near a gap-closing point, the lattice reduces to a two-band (1+1)-dimensional Dirac problem whose mass term changes sign across a domain wall \cite{SuSchriefferHeeger1979,SuSchriefferHeeger1980,RiceMele1982}.
Below we sketch this reduction and provide the analytic framework that guides our engineered domain-wall design.

\subsection{SSH to Dirac reduction\label{subsec:ssh_to_dirac}}
We begin with the two-site SSH tight-binding Bloch Hamiltonian with nearest-neighbour hoppings $t_1$ and $t_2$.
In the A/B sublattice basis with period $a$, it reads
\begin{equation}
\mathcal{H}(k)=\begin{pmatrix}
0 & -t_1-t_2e^{ika} \\[4pt]
-t_1-t_2e^{-ika} & 0
\end{pmatrix},
\label{eq:SSH_Hamiltonian}
\end{equation}
which can be expressed in terms of Pauli matrices as
\begin{equation}
\mathcal{H}(k)=-(t_1+t_2\cos(ka))\sigma_x+t_2\sin(ka)\sigma_y.
\label{eq:Pauli_components}
\end{equation}

To understand the gap physics, we perform low-energy linearizations at two high-symmetry points: the Brillouin-zone boundary $k=\pi/a$ and the zone center $k=0$.

Expanding about $k=\pi/a$ by setting $k=\pi/a+q$ with $|qa|\ll1$ gives to leading order
\begin{equation}
\mathcal{H}_{\pi}(q)\simeq-(t_1-t_2)\sigma_x-t_2qa\sigma_y.
\label{eq:H_pi_pre}
\end{equation}
The linearized Bloch Hamiltonian near $k=\pi/a$ thus becomes
\begin{equation}
\mathcal{H}_{\pi}(q)\simeq\-\Delta_{\pi}\sigma_x-vq\sigma_y,
\label{eq:H_pi}
\end{equation}
where we identify the effective mass term $\Delta_{\pi}=t_1-t_2$ and the velocity $v=t_2a$.
Transforming to real space via $q\mapsto-i\partial_x$, where we set $\hbar=1$, yields the effective Dirac operator
\begin{equation}
\mathcal{H}_{\pi}(x)\simeq-\Delta_{\pi}\sigma_x+iv\sigma_y\partial_x.
\label{eq:Dirac_pi}
\end{equation}

Performing the analogous expansion about the zone center $k=0$ gives a different effective mass term.
Setting $k=0+q$ with $|qa|\ll1$, we obtain
\begin{equation}
\mathcal{H}_{0}(q)\simeq-\Delta_{0}\sigma_x+vq\sigma_y,
\label{eq:H_0}
\end{equation}
where now $\Delta_0=t_1+t_2$ serves as the mass term and $v=t_2a$ as the velocity.
The corresponding real-space form is
\begin{equation}
\mathcal{H}_{0}(x)\simeq-\Delta_{0}\sigma_x-iv\sigma_y\partial_x.
\label{eq:Dirac_0}
\end{equation}

\subsection{Topological soliton as a Dirac zero mode\label{subsec:topological_solition}}
Having identified the low-energy Dirac forms and the effective masses $\Delta_{\pi}$ and $\Delta_0$, we now consider the situation where the mass varies slowly in space.
When $\Delta(x)$ changes sign across a spatial interface, the Dirac theory predicts a localized midgap bound state with zero energy: a Dirac zero mode.
This zero mode is a topological soliton whose existence is guaranteed by the change of a bulk topological invariant across the interface and does not depend on the detailed shape of the domain wall.

We first consider the Dirac Hamiltonian around $k=\pi/a$, Eq.~\eqref{eq:Dirac_pi}, with a spatially varying mass $\Delta(x)$.
The corresponding topological soliton solution takes the form
\begin{equation}
\psi^{\pi}_{\mathrm{soliton}}(x)\sim\begin{pmatrix}0\\1\end{pmatrix}\exp\left(-\frac{1}{v}\int^x\Delta(s)ds\right),
\label{eq:pi_soliton}
\end{equation}
up to a normalization factor.
Direct substitution shows that
\begin{equation}
\mathcal{H}_{\pi}(x)\psi^{\pi}_{\mathrm{soliton}}(x)=(\Delta(x)-\Delta(x))\exp\left(-\frac{1}{v}\int^x\Delta(s)ds\right)=0,
\end{equation}
confirming that Eq.~\eqref{eq:pi_soliton} is indeed the zero-energy wavefunction of the soliton.

Now assume that $\Delta(x)$ asymptotes to opposite constants at the two infinities as
\begin{equation}
\Delta(x)\to\begin{cases}
-\Delta_0, & x\to-\infty, \\
+\Delta_0, & x\to+\infty,
\end{cases}
\label{eq:jr_asympt}
\end{equation}
with $\Delta_0>0$.
For large $|x|$ the integrals behave as 
\begin{equation}
\int^x\Delta(s)ds\to\begin{cases}
-\Delta_0x, & x\to-\infty, \\
+\Delta_0x, & x\to+\infty.
\end{cases}
\end{equation}
Therefore the soliton's wavefunction decays exponentially as
\begin{equation}
\psi^{\pi}_{\mathrm{soliton}}(x)\sim\begin{pmatrix}0\\1\end{pmatrix}\exp\left(-\frac{\Delta_0|x|}{v}\right),\quad x\to\pm\infty,
\end{equation}
and we can define its localization length \cite{JackiwRebbi1976}
\begin{equation}
\xi_{\rm loc}\sim v/\Delta_0.
\end{equation}

For the Dirac Hamiltonian around $k=0$, Eq.~\eqref{eq:Dirac_0}, a similar topological soliton is given by
\begin{equation}
\psi^{0}_{\mathrm{soliton}}(x)\sim\begin{pmatrix}1\\0\end{pmatrix}\exp\left(-\frac{1}{v}\int^x\Delta(s)ds\right),
\label{eq:0_soliton}
\end{equation}
with the same localization length $\xi_{\rm loc}$.

In summary, whenever the effective Dirac mass $\Delta(x)$ changes sign across an interface, the low-energy Dirac description admits a single normalizable topological soliton localized at the domain wall.
This continuum theory of a domain wall in the Dirac mass provides the fundamental explanation for the midgap edge states observed in our bead-on-string chain.

\subsection{Designing the mechanical domain wall}
The above discussion clearly prescribes how to engineer a domain wall in the bead-on-string chain: assemble a finite left region whose intra-cell geometry yields an effective Dirac mass $\Delta(x)>0$, and a right region with the reversed intra-cell ordering so that $\Delta(x)<0$.
Stitching these two patterns together creates an interface where $\Delta(x)$ flips sign and traps a topological soliton.

In practice, we implement this by concatenating two dimerized sequences with opposite ordering of light and heavy beads.
We set the total chain length to $L=\SI{160}{cm}$.
Light beads of mass $m_{\mathrm{A}}=\SI{0.114}{g}$ are placed at positions $\SIlist{10;50;100;140}{cm}$, and heavy beads of mass $m_{\mathrm{B}}=\SI{0.282}{g}$ are placed at positions $\SIlist{20;60;90;130}{cm}$.
This configuration creates dimerized pairs $(\SI{10}{cm},\SI{20}{cm})$ and $(\SI{50}{cm},\SI{60}{cm})$ on the left side of the chain.
On the right side, the dimers are reversed to $(\SI{100}{cm},\SI{90}{cm})$ and $(\SI{140}{cm},\SI{130}{cm})$.
As a result, the effective Dirac mass $\Delta_{\pi}=t_1-t_2$ flips sign at the midpoint around $\SI{75}{cm}$, realizing the designed domain wall.

\begin{figure}
\includegraphics[width=\linewidth]{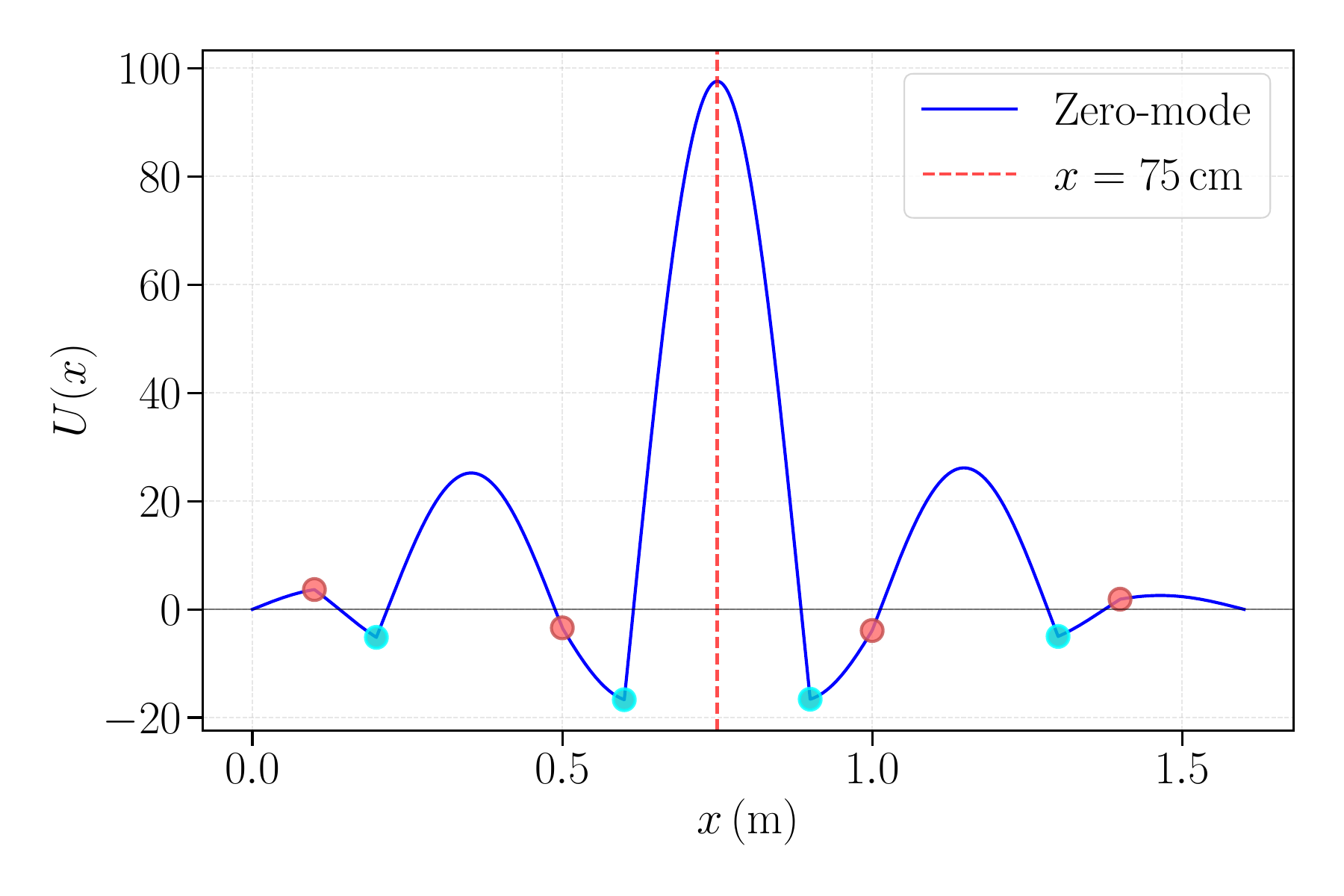}
\caption{Engineered domain-wall localized mode.
The chain has total length $L=\SI{160}{cm}$ with light beads $m_{\mathrm{A}}=\SI{0.114}{g}$ at positions \SIlist{10;50;100;140}{cm} and heavy beads $m_{\mathrm{B}}=\SI{0.282}{g}$ at positions \SIlist{20;60;90;130}{cm}.
This configuration produces dimerized pairs $(\SI{10}{cm},\SI{20}{cm})$ and $(\SI{50}{cm},\SI{60}{cm})$ on the left and reversed dimers $(\SI{100}{cm},\SI{90}{cm})$ and $(\SI{140}{cm},\SI{130}{cm})$ on the right, so that the effective mass $\Delta_{\pi}=t_1-t_2$ changes sign at the midpoint of $\SI{75}{cm}$, which coincides with the center of the trapped midgap topological soliton.
In the plot, the blue solid line shows the profile of the topological soliton, the red dashed line marks the domain-wall position at $x=\SI{75}{cm}$, red circles denote light beads with $m_{\mathrm{A}}=\SI{0.114}{g}$, and cyan circles denote heavy beads with $m_{\mathrm{B}}=\SI{0.282}{g}$.}
\label{fig:SSH_simulate}
\end{figure}

The computed mode in Fig.~\ref{fig:SSH_simulate} corresponds precisely to the midgap topological soliton localized at this engineered interface.
This agreement supports the validity of the two-band Dirac reduction in the parameter regimes we study.

\subsection{Further discussions\label{subsec:discussion}}
To conclude our analysis, we offer some additional remarks on the interpretation of edge states and their topological character.
The distinction between robust edge states in the first and third gaps versus fragile states in the second gap can be understood in terms of the effective Dirac masses derived from the SSH reduction.
In the uniform chain, where onsite energies are equal by symmetry, the low-energy expansions around $k=\pi/a$ and $k=0$ yield two distinct effective mass parameters: $\Delta_{\pi}=t_1-t_2$ and $\Delta_0=t_1+t_2$.
Crucially, the termination of the chain naturally creates a domain wall in $\Delta_{\pi}$, mimicking the situation in the SSH model where the edge site couples differently to the vacuum.
This sign reversal of $\Delta_{\pi}$ at the boundary induces topological edge states via the (1+1)-dimensional Dirac mass-domain mechanism \cite{JackiwRebbi1976}.
In contrast, no analogous sign change occurs for the positive definite $\Delta_0$, which explains why edge states associated with band edges near $k=\pi/a$ are robust while those associated with $k=0$ remain termination-sensitive.

For the dimerized chain, the situation is more subtle because the two beads in a unit cell have different masses $m_{\mathrm{A}}\neq m_{\mathrm{B}}$, introducing unequal onsite energies.
A strict mapping to the SSH model, which assumes equal onsite energies, is therefore not exact.
Nevertheless, the qualitative pattern persists: edge states in the first and third extra-gaps remain robust against local boundary perturbations, while the state in the second extra-gap does not.
This suggests that the underlying Dirac picture still provides a useful conceptual guide even when onsite differences are present.

The observed contrast between robust and fragile edge states bears a phenomenological resemblance to the classic distinction between Shockley and Tamm surface states in electronic systems \cite{Shockley1939,Tamm1932}.
Shockley states are topologically mandated by the bulk band structure and are robust to local boundary details, analogous to our robust edge modes in the first and third gaps.
Tamm states, on the other hand, arise from strong surface perturbations and are sensitive to termination details, much like the fragile mode we find in the second gap.
While our mechanical system does not possess the same quantum numbers as electronic crystals, this analogy offers a helpful intuitive picture for understanding why certain edge modes survive under parameter changes while others do not.

\section{Conclusion\label{sec:conclusion}}
We demonstrate that a one-dimensional bead-on-string chain provides a clear platform for studying band modulations and topological transitions.
Using an exact transfer-matrix formulation, we characterize how bead mass and spacing variations tune band gaps and edge states.

Our analysis reveals a systematic pattern: in both uniform and dimerized configurations, robust edge modes emerge in the first and third gaps, while the second gap hosts only termination-sensitive states.
This distinction finds a natural interpretation within the SSH-like Dirac reduction framework.
The successful engineering of a domain wall that traps a midgap topological soliton further validates this interpretation and underscores the universality of such topological phenomena across different physical platforms.

\section*{Acknowledgments\label{sec:acknowledgements}}
We thank Prof. Kun Xun for introducing the bead-on-string chain system into the undergraduate physics experiment course, which inspired this work.
This work is supported by the National Key Research and Development Program of China under Grant No. 2024YFA1209202(C.L.).

\bibliographystyle{apsrev4-2}
\bibliography{references}

\end{document}